\documentclass[journal=langd5,manuscript=article]{achemso}
\usepackage[version=3]{mhchem} 
\usepackage{natbib}
\usepackage{url}
\usepackage{appendix}
\usepackage{subcaption}

\usepackage{epstopdf}
\usepackage{mathtools}
\usepackage{textgreek}

\usepackage{multirow}
\usepackage{multicol}
\usepackage{graphicx}

\usepackage{caption}

\usepackage{amssymb}
\usepackage{float}
\usepackage[export]{adjustbox}
\usepackage{slashbox}

\usepackage{xcolor}

\newcommand{\ks}{\textcolor{black}} 

\author{A. Hari Govindha}
\affiliation{Department of Mechanical and Aerospace Engineering, Indian Institute of Technology Hyderabad, Kandi - 502284, Telangana, India}

\author{Sayak Banerjee}
\affiliation{Department of Mechanical and Aerospace Engineering, Indian Institute of Technology Hyderabad, Kandi - 502284, Telangana, India}
\author{Saravanan Balusamy}
\affiliation{Department of Mechanical and Aerospace Engineering, Indian Institute of Technology Hyderabad, Kandi - 502284, Telangana, India}
\author{Kirti Chandra Sahu}
\affiliation{Department of Chemical Engineering, Indian Institute of Technology Hyderabad, Kandi - 502284, Telangana, India}
\email{ksahu@che.iith.ac.in}

\title{Evaporation Dynamics of Completely Pinned and Partially Pinned Sessile Droplets in Multi-Droplet Configurations}

\begin{document}

\maketitle
\begin{abstract}
The evaporation of sessile droplets placed in close proximity is influenced by complex vapor–vapor interactions, producing a shielding effect that can significantly extend droplet lifetimes. This study presents a systematic experimental investigation of evaporation dynamics in multi-droplet configurations under ambient conditions, comparing completely pinned and partially pinned contact line modes. Completely pinned droplets are generated by introducing alumina nanoparticles, while partially pinned droplets consist of pure water. An isolated droplet is compared with arrays of two, three, and five droplets, each arranged at a fixed spacing ratio. High-speed shadowgraphy is used to measure droplet height, contact angle, volume, and lifetime. Results show that, although completely pinned droplets evaporate faster in absolute terms due to a constant contact radius, they experience a more pronounced relative lifetime increase from vapor shielding than partially pinned droplets. In five-droplet configurations, lifetimes increase by up to 89\% and 124\% compared to isolated droplets for partially pinned and completely pinned modes, respectively. A theoretical model incorporating evaporative cooling predicts central droplet lifetimes with good agreement. These findings underscore the coupled influence of contact line mobility and droplet proximity on evaporation rates.
\end{abstract} 


\section{Introduction}
\label{sec:intro}

Understanding the evaporation and wetting dynamics of sessile droplets is crucial for elucidating various natural phenomena and serves as the foundation for numerous technological applications \cite{Gurrala2021}. These include inkjet printing for producing femtoliter droplets \cite{de2004inkjet,tekin2004ink,soltman2008inkjet,park2006control}, modeling disease transmission through respiratory droplets \cite{balusamy2021lifetime,bhardwaj2020likelihood}, microarray fabrication for controlling DNA morphology, sensitivity, and distribution in biosensing \cite{dugas2005droplet,lee2006electrohydrodynamic}, coating technologies in the pharmaceutical and ceramic industries \cite{kim2016controlled,pahlavan2021evaporation}, architectural painting processes \cite{brinckmann2011experimental}, spray and hotspot cooling for localized thermal management \cite{kim2007spray,ruan2019effects,cheng2010active}, and microfluidic applications \cite{deegan1997capillary,kumar2020wetting}, to name a few. While extensive research has been devoted to the evaporation of isolated sessile droplets, many real-world applications involve multiple droplets arranged in close proximity, either in structured arrays or random distributions on a surface. In such configurations, vapor emitted from neighboring droplets raises the local humidity and suppresses the evaporation flux of individual droplets. This phenomenon is known as the shielding effect \cite{schofield2020shielding}. Consequently, a comprehensive understanding of the collective evaporation behavior in droplet ensembles is essential for accurately predicting and controlling evaporation dynamics in practical applications. In the following, we provide a brief overview of key studies on sessile droplet evaporation, covering both isolated single-droplet systems and multi-droplet configurations.

The evaporation behavior of an isolated sessile droplet has been extensively studied under various conditions \cite{mondal2023physics,katre2020evaporation}. \citet{birdi1989study} reported that sessile water droplets on hydrophilic substrates exhibit a nearly constant evaporation rate. Two distinct evaporation modes, namely constant contact radius (CCR) and constant contact angle (CCA), were identified by \citet{picknett1977evaporation}. Furthermore, it was observed that the transition to the CCA mode occurs after the initial CCR stage, typically when the droplet reaches the receding contact angle. Additionally, \citet{bourges1995influence} categorized the evaporation process into four distinct stages during the evaporation of n-decane and pure water droplets on hydrophilic substrates with varying roughness. The influence of surface wettability on evaporation dynamics was explored by \citet{shin2009evaporating}, who observed distinct contact line behaviors for hydrophilic, hydrophobic, and superhydrophobic surfaces. \citet{yu2012experimental} investigated droplet evaporation on Teflon and polydimethylsiloxane (PDMS) substrates and identified the presence of CCR, CCA, and mixed modes during evaporation. \citet{girard2010infrared, girard2008effect} analyzed the temperature distribution between the apex and contact line of evaporating water droplets and concluded that Marangoni flows had a negligible influence on the evaporation rate. \citet{gurrala2019evaporation} conducted a combined experimental and theoretical study on the evaporation of pure and binary sessile droplets. They proposed a theoretical model that incorporates diffusion, natural convection, and passive vapour transport driven by air movement. Their results revealed a non-monotonic variation in droplet lifetime with increasing ethanol concentration at elevated substrate temperatures. In addition, numerous studies have examined the influence of substrate thermal conductivity \cite{dunn2009strong,sobac2012thermal,ristenpart2007influence}, thermal and solutal Marangoni convection \cite{van2021marangoni,kim2016controlled}, and substrate characteristics such as roughness, geometry, and inclination \cite{pittoni2014uniqueness,gurrala2022evaporation,katre2022experimental} on the evaporation dynamics of sessile droplets. These investigations, employing a combination of analytical, numerical, and experimental methods, have significantly advanced the current understanding of droplet evaporation under diverse physical conditions.

In isolated droplet configurations, the evaporation dynamics of droplets laden with nanoparticles have also been investigated to understand the influence of suspended particles on contact line behavior and deposit formation \cite{orejon2011stick,katre2021evaporation}. \citet{orejon2011stick} investigated the three-phase contact line dynamics of pure water and ethanol droplets on substrates with varying degrees of hydrophobicity, showing that higher hydrophobicity promotes contact line depinning. \citet{hari2022counter} performed experiments with ethanol droplets containing alumina and copper nanoparticles, concluding that contact line motion plays a more dominant role in governing evaporation rates than the thermal conductivity of the nanoparticles. \citet{patil2016effects} examined the combined effects of substrate temperature, particle concentration, and surface wettability on droplet evaporation and deposition, observing ring-shaped deposits forming in the inner region at elevated temperatures.
Distinct deposition patterns—such as uniform, dual-ring, and stick-slip—were observed by \citet{parsa2015effect} in water droplets containing copper oxide nanoparticles, with the variations attributed to changes in the balance between capillary forces and Marangoni flows across different substrate temperatures. \citet{moffat2009effect} reported that increasing the concentration of TiO$_2$ nanoparticles in ethanol droplets enhances stick-slip motion on PTFE-coated silicon wafers. \citet{kovalchuk2014evaporation} found that increasing nanoparticle concentration leads to a higher overall diffusive evaporation rate. The evaporation rate was also shown to be strongly dependent on both the type and concentration of nanoparticles, as demonstrated by \citet{moghiman2013influence}. To improve deposit uniformity, \citet{yunker2011suppression} employed ellipsoidal particles, which effectively suppressed the coffee-ring effect. Additionally, \citet{vafaei2009effect} reported that both the contact angle and particle size increase with rising nanoparticle concentration.

Next, we discuss the literature on the complex interactions that govern evaporation dynamics in multi-droplet configurations \cite{wray2024high,Hari2024,kavuri2025jfm}. \citet{shaikeea2016insight} examined a two-droplet system on a hydrophobic surface and observed a reduction in evaporation rates along with asymmetric contact line dynamics between the left and right droplets during the early stages of evaporation. \citet{carrier2016evaporation} conducted both experimental and theoretical investigations comparing isolated and multiple droplet systems. They found that the presence of neighboring droplets leads to reduced evaporation rates due to local atmospheric saturation, a phenomenon analogous to the shielding effect observed in dissolution processes. In a related study, \citet{laghezza2016collective} demonstrated increased dissolution times for alcohol droplets submerged in water when compared to isolated droplets, highlighting the complexity of vapor field interactions in multi-droplet systems. \citet{pradhan2015deposition} experimentally analyzed internal flows and deposition patterns in two-droplet systems and found that, due to non-uniform evaporation rates between the right and left contact lines of a droplet, particle deposition near the contact lines adjacent to the droplets is reduced. Universal evaporation behaviors in arrays of three droplets and $5 \times 5$ two-dimensional (2D) droplet systems were reported by \citet{hatte2019universal} and \citet{pandey2020cooperative}, who employed a scaling approach to relate droplet lifetime to the $L/d$ ratio, where $d$ denotes the equatorial droplet diameter and $L$ is the distance between the side and central droplets. \citet{khilifi2019study} investigated the evaporation dynamics of a seven-droplet configuration on sapphire substrates and found that droplet lifetimes decreased with increasing inter-droplet distance. This trend was consistent with the competitive diffusion-limited evaporation model proposed by \citet{wray2020competitive}. In their study, multiple droplets were simultaneously deposited on the same substrate under identical conditions, allowing for a direct comparison between the central droplet and an isolated single droplet. The results highlighted the influence of neighboring droplets in slowing down the evaporation process, particularly for the centrally located droplet, due to vapor field interactions. Analytical models developed by \citet{tonini2022analytical} for two- and three-droplet systems on both hydrophilic and hydrophobic substrates indicated that evaporation rate reductions were more pronounced on hydrophobic surfaces for a given $L/d$ ratio. Recently, \citet{wray2024high} studied an arbitrary array of potential non-circular sources and derived an asymptotic solution that provides accurate and rapid results for a wide range of problems. The numerical solutions were also validated against experimental data. \citet{masoud2021evaporation} extended this modeling approach to predict diffusive evaporation rates for droplets of varying sizes and contact angles. Recent advances in volume measurement, such as interferometry \cite{edwards2021interferometric}, have enabled precise quantification of droplet volumes, overcoming limitations of conventional imaging methods. Using a pattern distortion technique, \citet{iqtidar2023drying} tracked the volume evolution of central and side droplets across various configurations and reported consistent drying behavior among them. Theoretical and numerical models developed by \citet{rehman2023effect} and \citet{chen2022predicting} further examined the evaporation dynamics of multiple droplets on hydrophilic and hydrophobic surfaces, respectively. \citet{Hari2024} carried out experimental investigations on multiple droplets evaporating at elevated substrate temperatures and various $L/d$ ratios on hydrophobic surfaces, revealing a power-law dependence of droplet lifetime on temperature and a reduction in the shielding effect due to enhanced natural convection. Their study also incorporated theoretical models that accounted for evaporative cooling. Additionally, \citet{kavuri2025evaporation} examined the evaporation of two droplets on superhydrophobic substrates at elevated temperatures, developing a theoretical framework that integrated diffusion, evaporative cooling, and natural convection.

As the above review suggests, while recent studies have examined droplet interactions in specific configurations, such as droplet pairs, linear arrays, and clusters, a comprehensive understanding of how geometric arrangement and contact line dynamics influence evaporation kinetics is still lacking. Key aspects such as the interplay between droplet proximity, the number of droplets, and contact line mobility (i.e., evaporation in completely pinned versus partially pinned droplets) remain inadequately addressed. In the present study, the completely pinned mode refers to droplets whose contact radius remains constant throughout their lifetime, whereas the partially pinned mode refers to droplets whose contact radius varies during evaporation. Moreover, most existing investigations primarily consider purely diffusive transport or idealized boundary conditions, often overlooking critical physical effects like evaporative cooling and natural convection, which can significantly influence evaporation even at room temperature. To address these gaps, the present study investigates the evaporation behavior of multiple sessile droplets under ambient conditions, focusing specifically on the role of contact line dynamics. Two distinct modes, namely completely pinned and partially pinned contact lines, are considered, each resulting in different levels of vapor shielding during evaporation. The completely pinned mode is achieved by introducing nanoparticles, which effectively anchor the contact line to the substrate. This study explores four droplet configurations: a single isolated droplet and arrays comprising two, three, and five droplets. We perform detailed experiments for each configuration to examine the evaporation process, estimating droplet lifetimes and tracking the temporal evolution of key parameters such as height, wetted diameter, contact angle, and volume. In addition, a theoretical model incorporating evaporative cooling effects is developed and employed to predict the lifetime of the central droplet within the array.

\section{Experimental Methodology}
\label{sec:expt}
\subsection{Experimental Setup}
 
\begin{figure}[h]
\centering
\includegraphics[width=0.9\textwidth]{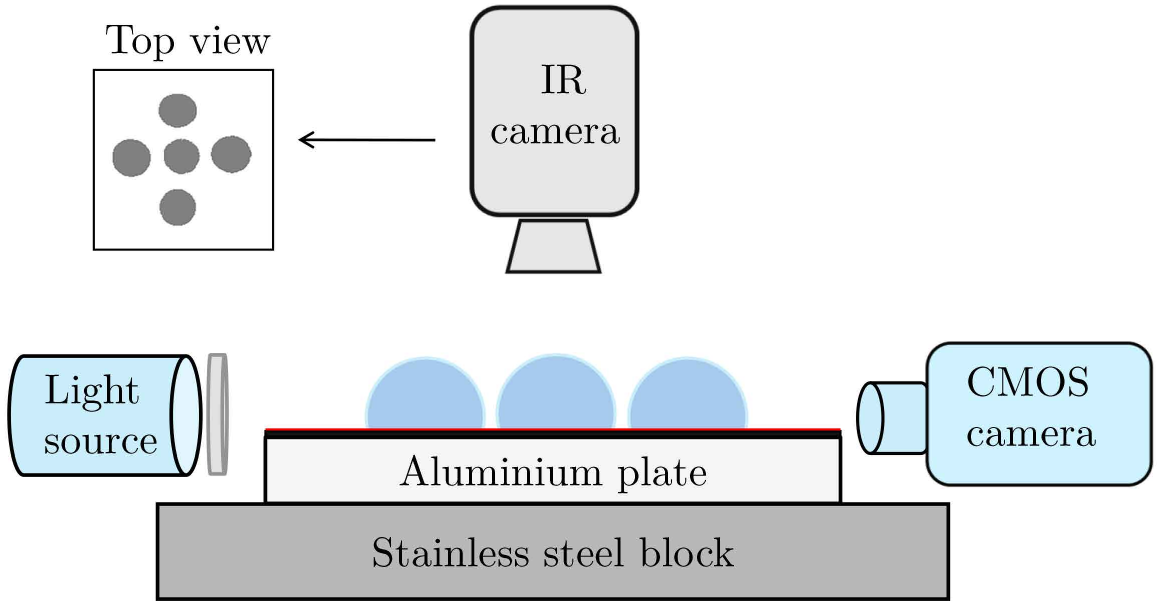}
\caption{Schematic diagram of the experimental setup. The side view of the droplets is captured using a CMOS camera in a shadowgraphy arrangement, while the top view is recorded by an infrared (IR) camera. Droplets are deposited on an aluminum substrate coated with a 100 $\mu$m thick polytetrafluoroethylene (PTFE) tape.} 
\label{fig:fig1}
\end{figure}

We experimentally investigate the evaporation dynamics of completely pinned versus partially pinned sessile droplets arranged in various configurations on a substrate, employing shadowgraphy and infrared (IR) imaging techniques. The completely pinned droplets are achieved by introducing nanoparticles, which anchor the contact line to the substrate. This approach enables a direct comparison between completely pinned and partially pinned evaporation modes in multi-droplet systems under identical ambient and geometric conditions. A schematic of the experimental setup is shown in Figure \ref{fig:fig1}(a). The setup comprises a multilayered metallic block, a motor-driven pump for dispensing droplets onto the substrate, and a proportional-integral-derivative (PID) controller to regulate the flow rate and volume of the dispensed droplets. Imaging is carried out using a complementary-metal-oxide-semiconductor (CMOS) camera (Make: Do3Think, Model: DS-CBY501E-H) and an infrared (IR) camera (Make: FLIR, Model: X6540sc). An LED light source, equipped with a diffuser sheet, ensures uniform illumination. The CMOS and IR cameras are used to capture the side and top views of the evaporating droplet, respectively. The entire assembly is housed inside a customized goniometer box (Make: Holmarc Opto-Mechatronics Pvt. Ltd.) to minimize external environmental disturbances. All experiments are conducted at a controlled room temperature of $(24 \pm 1)^\circ$C and relative humidity $RH=(55 \pm 5)\%$. The RH is monitored using a hygrometer (Make: HTC, Model: 288-ATH) installed within the goniometer enclosure.

The multilayered metallic block consists of a stainless steel base topped with an aluminum plate measuring 100 mm × 80 mm × 15 mm. A thin coat of black paint is applied to the aluminum surface to minimize reflections in the infrared (IR) images and ensure accurate temperature measurements. This block is mounted on a movable stage (Figure \ref{fig:fig1}(b)) that can traverse in both longitudinal and lateral directions using a lead screw mechanism. A complementary-metal-oxide-semiconductor (CMOS) camera with a spatial resolution of 2592 × 1944 pixels captures the side view of the droplet at a rate of one frame per second (fps). These side-view images provide key droplet parameters, including the wetted diameter ($D$), height ($h$), contact angle ($\theta$), and volume ($V$). Top views of the droplets are acquired using an IR camera with a resolution of 640 × 512 pixels at 1 fps, operating in the spectral range of 3–5 μm. This allows for simultaneous monitoring of the entire droplet array during evaporation. A 100 $\mu$m thick polytetrafluoroethylene (PTFE) tape is affixed to the aluminum plate and used as the substrate. Prior to use, the PTFE surface is cleaned with isopropanol, dried with compressed air, and then securely pasted onto the aluminum plate. 

\begin{figure}[htbp]
\centering
\includegraphics[width=0.5\textwidth]{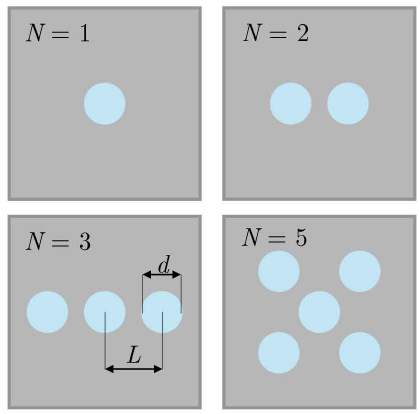}
\caption{Various droplet configurations considered in the present study. Here, $N$ denotes the number of droplets arranged in each configuration, $d$ is the equatorial diameter of a droplet, and $L$ represents the center-to-center distance between the central and side droplets.} 
\label{fig:fig2}
\end{figure}

Nanoparticle-laden solutions are prepared by dispersing 25 nm Al$_2$O$_3$ nanoparticles (Sisco Research Laboratories Pvt. Ltd.) in deionized water on a weight percentage (wt.\%) basis. The solution is then sonicated in an ultrasonic unit (Make: BRANSON, Model: CPX1800H-E) to ensure uniform dispersion of nanoparticles. A 50 μL chromatography syringe (Make: Unitek Scientific Corporation, piston size: 1.08 mm) fitted with a 21G needle (inner orifice diameter: 0.514 mm) is connected to a motorized pump to control the flow rate and dispense droplets of consistent volume. Following the deposition of each droplet, the substrate is translated by a known distance in the longitudinal or lateral direction using the lead screw mechanism, depending on the required configuration. The droplet arrangements range from a single isolated droplet to a five-droplet array, as illustrated in Figure \ref{fig:fig2}. The spacing between the central and side droplets is denoted by $L$, and $d$ represents the equatorial diameter of a droplet. All experiments are conducted at a fixed configuration ratio of $L/d = 1.2$. For isolated droplets, it is well established that pinned droplets evaporate faster than partially pinned ones due to differences in contact line dynamics. In a multi-droplet system, however, evaporation rates are jointly influenced by droplet proximity and contact line behavior. To maximize inter-droplet interactions while avoiding frequent coalescence, which becomes dominant at smaller $L/d$ values and compromises repeatability, we selected $L/d = 1.2$. Furthermore, the influence of $L/d$ on evaporation in partially pinned droplet configurations has been systematically examined in \citet{Hari2024}, and their findings can be reasonably extrapolated to the present case of completely pinned droplets under a fixed $L/d$.

The volume of each generated droplet was approximately $1.4 \pm 0.1~\mu$l. In all experiments, time $t = 0$ corresponds to the instant the droplet first contacts the substrate. Even in the five-droplet configuration, the total time required to place all droplets was only about 4\% of the total lifetime of the droplets, allowing us to neglect this delay in the analysis. To maintain consistent surface conditions, the PTFE tape was replaced after each experiment. For each set of parameters, experiments were repeated at least three times to ensure reproducibility.

\subsection{Post-processing}

Post-processing of the side-view images captured by the CMOS camera was carried out using a custom program developed in the \textsc{Matlab}$^{\circledR}$ environment to extract the droplet profiles. The procedure began with the application of a median filter to remove random noise, followed by unsharp masking to enhance image sharpness and accentuate gradients. The enhanced image was then converted to a binary format using an appropriate threshold to clearly delineate the droplet boundary from the background. Internal holes within the droplet silhouette were filled, and any reflected portion of the droplet was removed. The droplet contour was then traced using a built-in \textsc{Matlab}$^{\circledR}$ function, enabling the extraction of geometric parameters such as height, wetted diameter, contact angle, and volume. The image processing approach employed in the present study is similar to the methodology described by \citet{gurrala2019evaporation} and is therefore not detailed further; readers are referred to our previous studies for more information.
 
\section{Results and Discussion}\label{sec:results}

\subsection{Droplet Lifetimes: Comparative Analysis}

We investigate the evaporation dynamics of sessile droplet arrays on a hydrophobic substrate in various configurations, using both pure water droplets and droplets containing 2.0 wt. \% Al$_2$O$_3$ nanoparticles. We observe that pure water droplets exhibit partially pinned contact-line behavior, whereas nanoparticle-laden droplets remain completely pinned. To assess the influence of nanoparticle loading on evaporation dynamics, additional experiments were conducted on a cellulose acetate substrate, which naturally pins the contact line without nanoparticles. The results (Supplementary Table S1) show that the lifetimes of droplets with and without nanoparticles differ by less than 4\%, indicating that nanoparticles serve primarily to pin the contact line without significantly affecting the evaporation rate. In all configurations considered, except for the single- and two-droplet cases, a central droplet is surrounded by side droplets positioned at a fixed distance. The spacing is characterized by the $L/d$ ratio (center-to-center distance $L$ to droplet diameter $d$), set to 1.2 for both completely pinned and partially pinned arrangements. This quantifies the spacing between droplets relative to their equatorial diameter. In the present study, pure water droplets evaporate in the partially pinned mode, whereas nanoparticle-laden droplets evaporate in the completely pinned mode. Table \ref{table:T1} lists the values of the droplet lifetime $(t_e)$ for the various configurations. For the three- and five-droplet cases, the value of $t_e$ corresponds to the central droplet, which evaporates more slowly than the surrounding droplets. For the two-droplet case, $t_e$ represents the time required for both droplets to evaporate completely. It can be seen that the largest deviation in measured lifetimes $\pm 5.6\%$ occurs in the partially pinned three-droplet configuration.

\begin{table}[]
\centering
\caption{Lifetime (in seconds) of the longest-lasting droplet in partially pinned and completely pinned in different configurations.}
\label{table:T1}
\hspace{2 mm}\\
\begin{tabular}{|l|ll|}
\hline
\multicolumn{1}{|c|}{\multirow{2}{*}{$N$}} & \multicolumn{2}{c|}{$t_e$ (s)}                      \\ \cline{2-3} 
\multicolumn{1}{|c|}{}                   & \multicolumn{1}{l|}{Partially pinned droplet}     & Completely pinned droplet      \\ \hline
1                                        & \multicolumn{1}{l|}{$1900\pm28$}  & $1500\pm49$  \\ \hline
2                                        & \multicolumn{1}{l|}{$2490\pm90$}  & $2119\pm28$  \\ \hline
3                                        & \multicolumn{1}{l|}{$2925\pm165$} & $2595\pm45$  \\ \hline
5                                        & \multicolumn{1}{l|}{$3600\pm120$} & $3360\pm120$ \\ \hline
\end{tabular}
\end{table}

\begin{figure}[h]
\centering
\hspace{0.5cm}{\large (a)} \hspace{7.1cm}{\large (b)} \\
\includegraphics[width=0.45\textwidth]{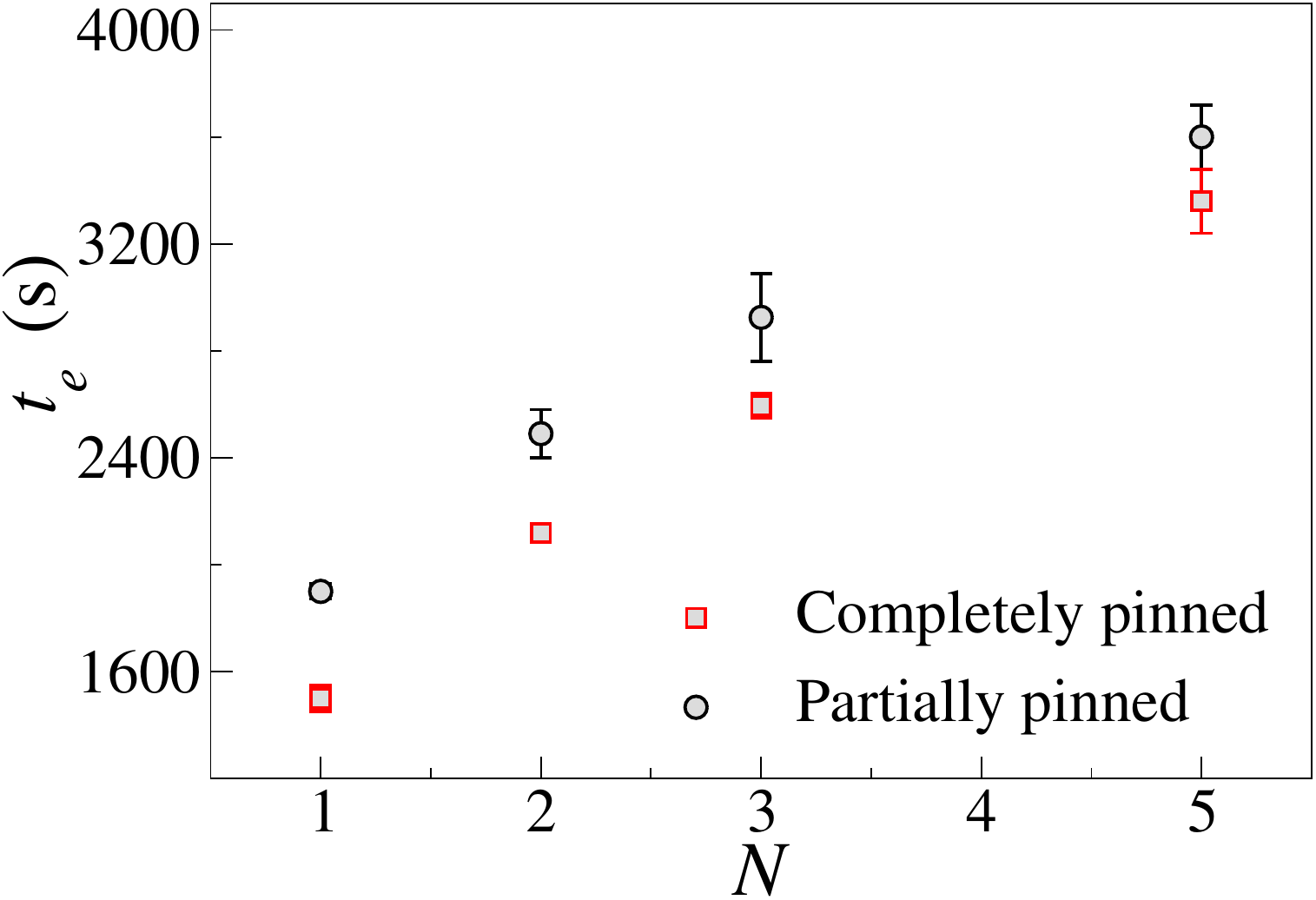} \hspace{2mm} \includegraphics[width=0.42\textwidth]{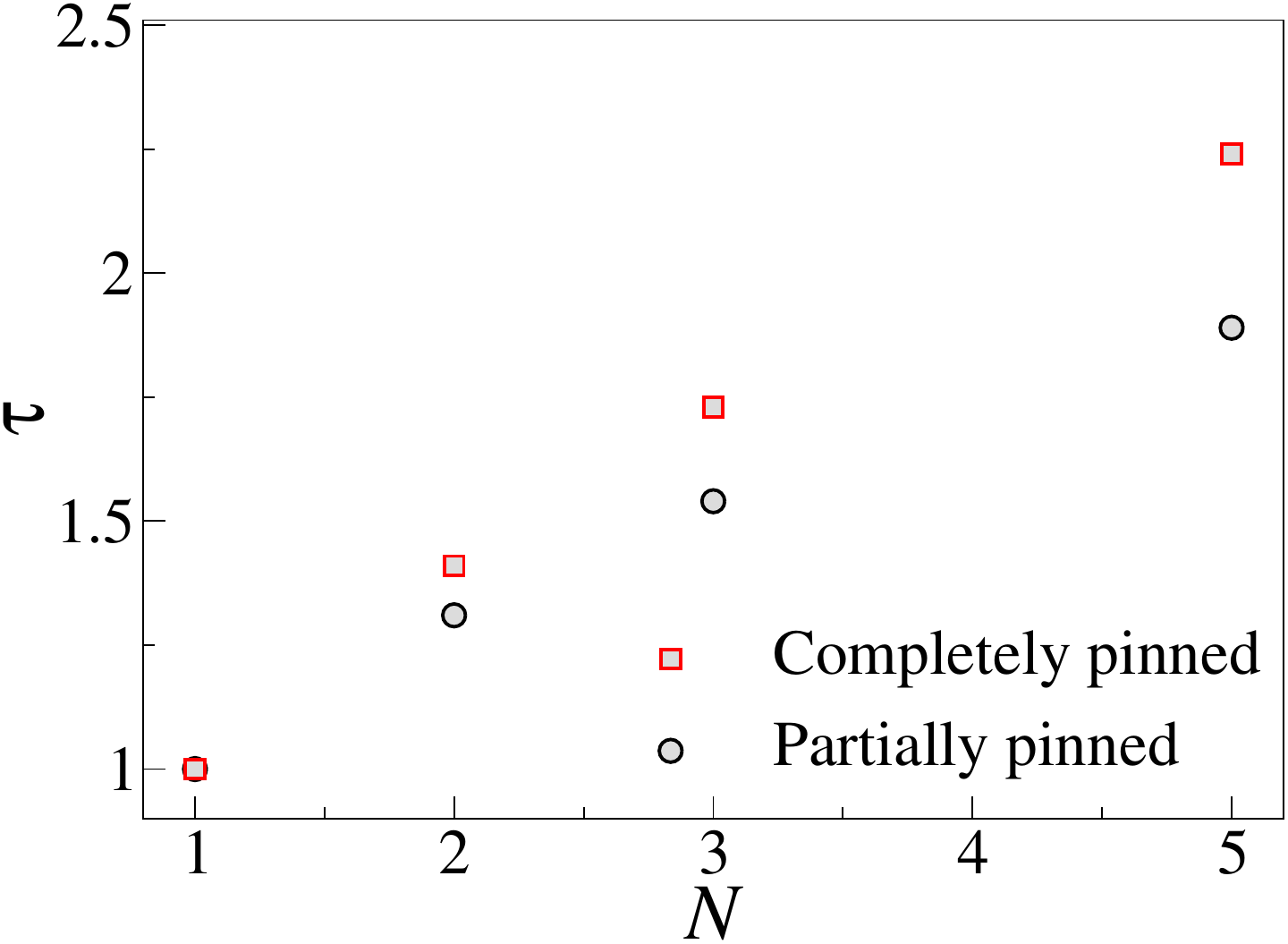}\\
\hspace{0.5cm}{\large (c)}\\
\hspace{-0.2cm}\includegraphics[width=0.45\textwidth]{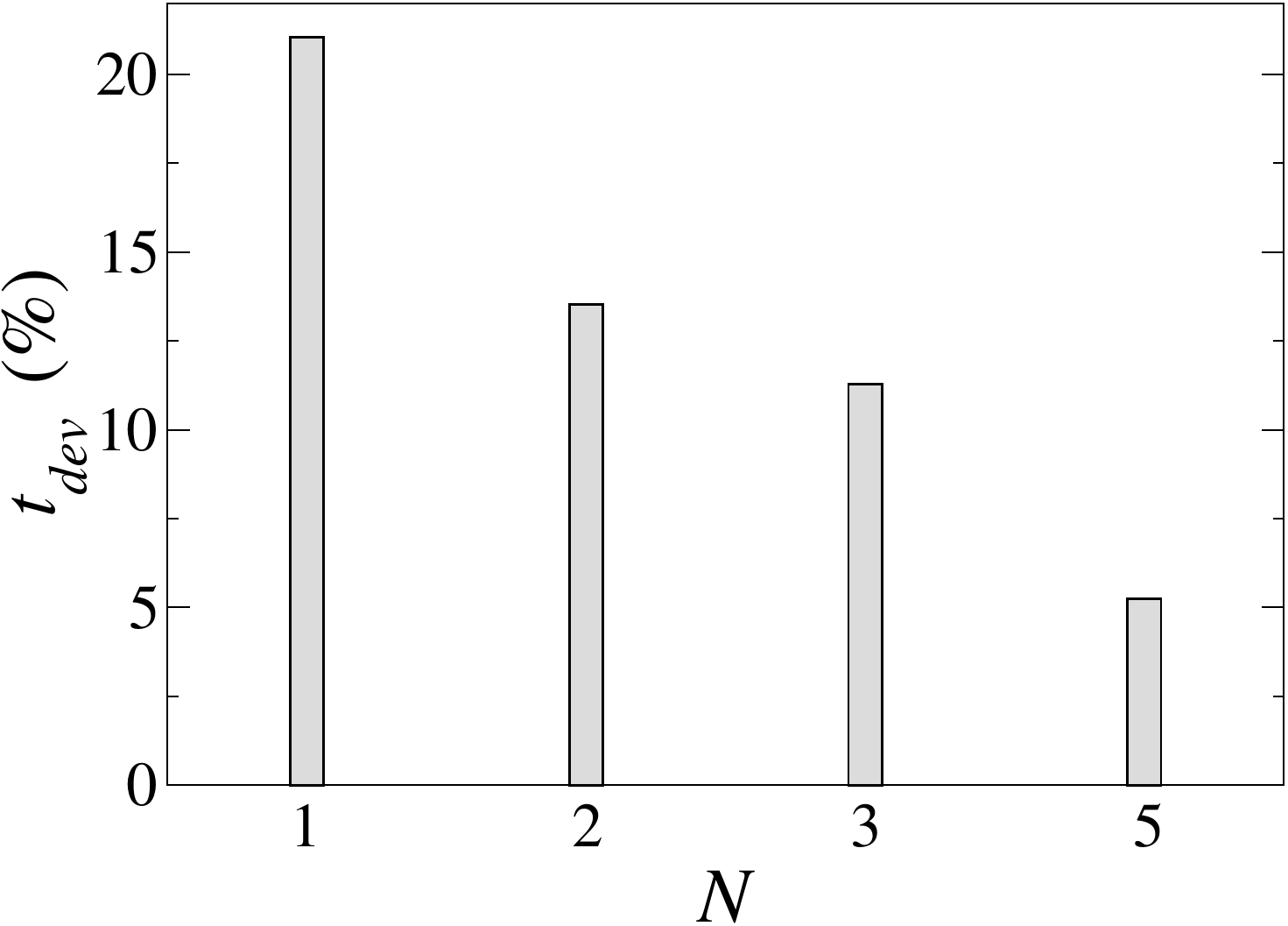} \hspace{2mm}\\
\caption{Variation of (a) droplet lifetimes (in seconds), (b) normalized droplet lifetime ($\tau$), and (c) relative deviation in lifetime ($t_{dev}$ in \%) between partially pinned and completely pinned droplets with $N$.}
\label{fig:fig3}
\end{figure}

The variation in droplet lifetimes with increasing numbers of droplets is shown in Figure~\ref{fig:fig3}(a) for both completely pinned and partially pinned cases. The lifetime increases with the number of droplets in the configuration for both modes. The best linear fits for lifetime versus droplet number are provided in Supplementary Table S2. This increase is attributed to the vapor shielding effect~\cite{chen2022predicting,schofield2020shielding}, arising from the diffusion-based interaction of vapor fields generated by all droplets in the array. As the number of droplets increases, the local vapor concentration around each droplet rises, reducing the concentration gradient driving evaporation and thereby lowering the evaporation rate. For all configurations studied, the lifetime of completely pinned droplets is consistently lower than that of their partially pinned counterparts. This is because, in the completely pinned mode, the contact radius and free surface area remain constant throughout evaporation, leading to a higher evaporation rate. The lifetime of an isolated partially pinned droplet is approximately 1900 seconds, with a maximum increase of 89\% observed for the five-droplet configuration. In comparison, completely pinned droplets have an isolated lifetime of about 1500~seconds, with a maximum increase of 124\% for the five-droplet arrangement. These results indicate that the relative influence of vapor shielding is greater for completely pinned droplets, even though their absolute evaporation rates remain higher than those of partially pinned droplets.

\begin{figure}[h]
\centering
\hspace{0.5cm}{\large (a)} \hspace{7.1cm}{\large (b)} \\
 \includegraphics[width=0.45\textwidth]{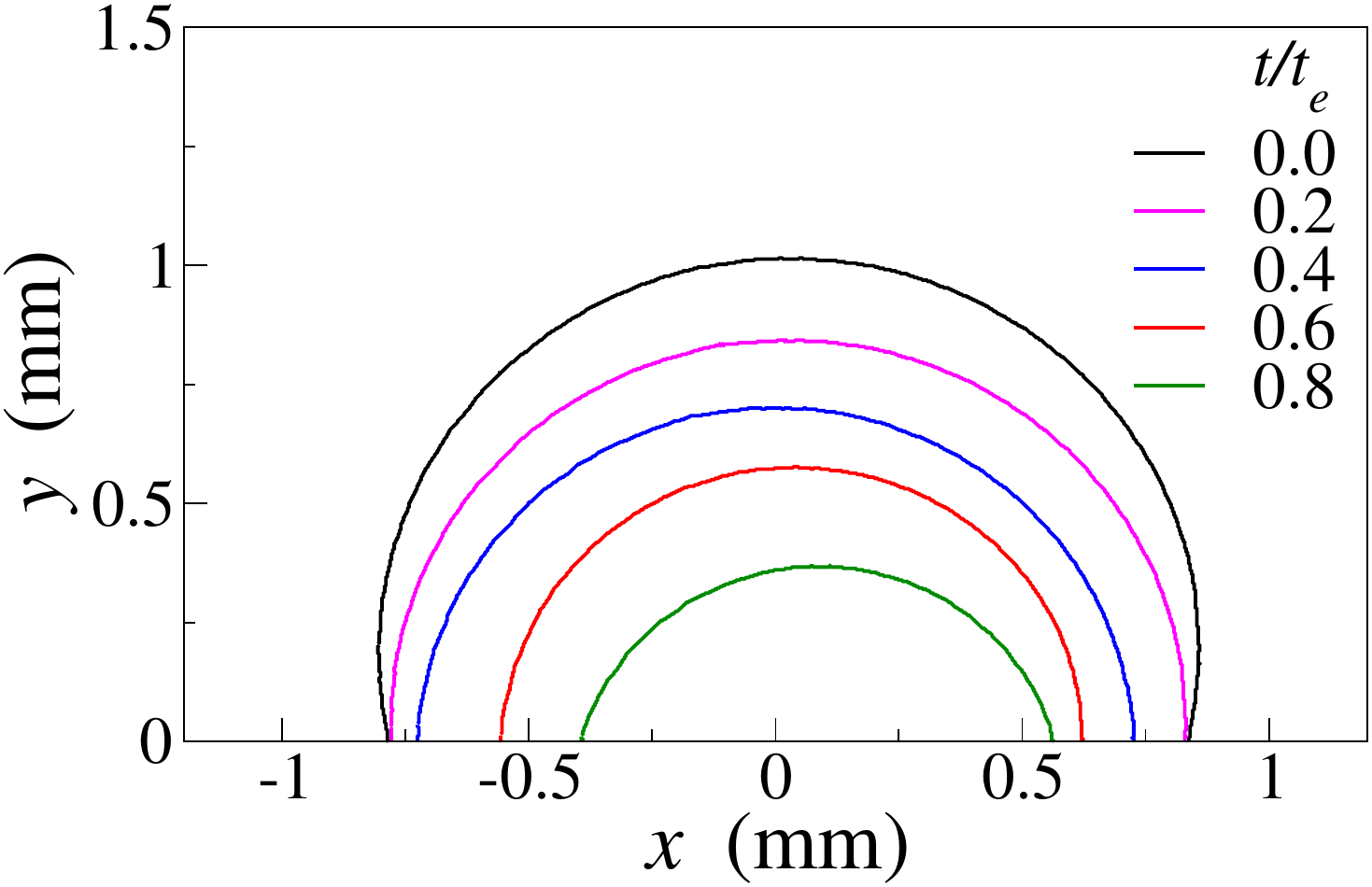} \hspace{2mm} \includegraphics[width=0.45\textwidth]{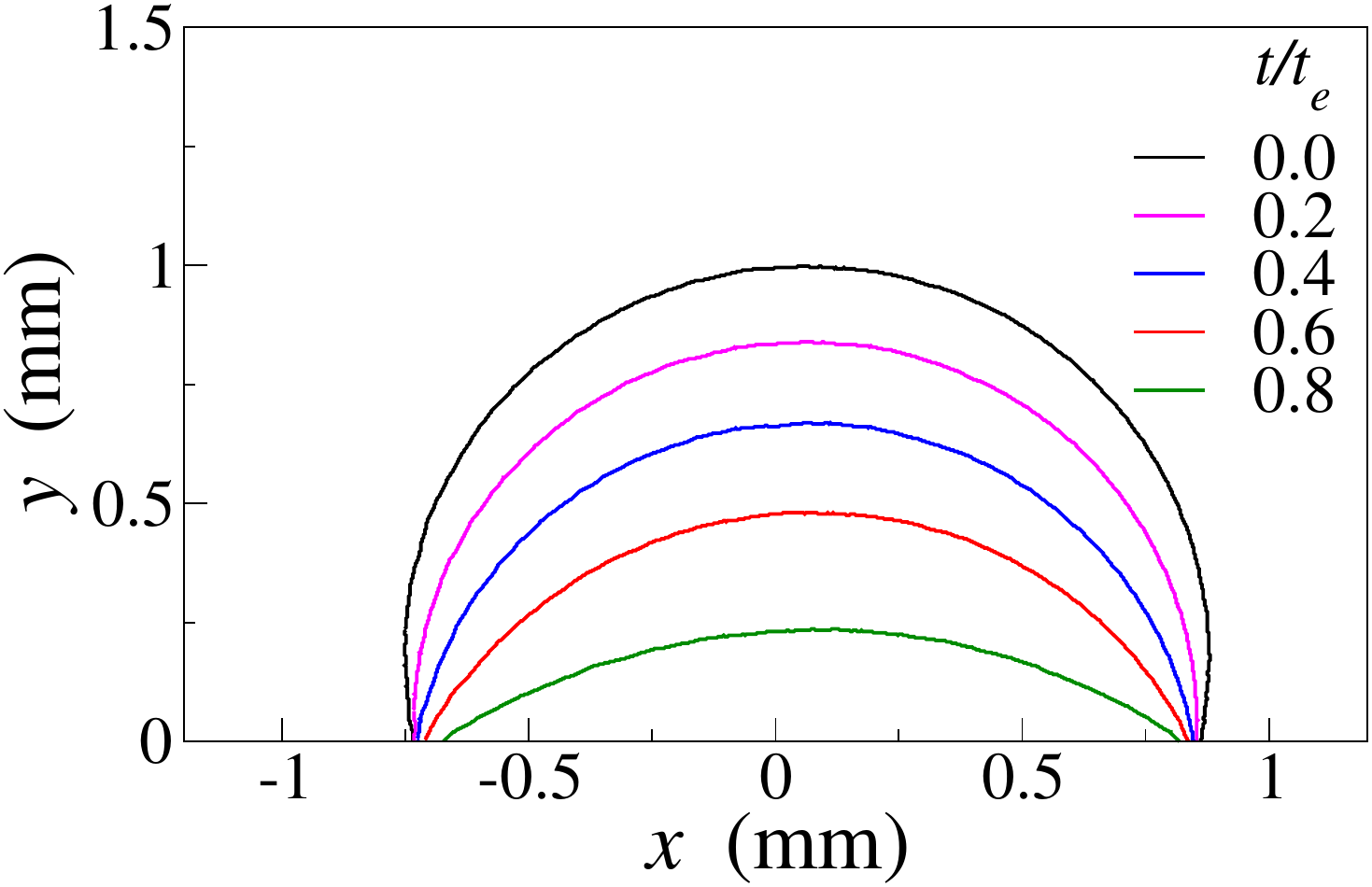}\\
\caption{Temporal evolution of droplet contours for the (a) partially pinned and (b) completely pinned configurations.}
\label{fig:fig4}
\end{figure}

The pinning or depinning of the contact line is illustrated in Figure~\ref{fig:fig4}, which presents the temporal evolution of the droplet side contour profiles superimposed at different dimensionless times ($t/t_e$). The $x$- and $y$-axes represent the droplet spread and height, respectively. Contours are shown from the initial state ($t/t_e = 0$) to 80\% of the droplet lifetime ($t/t_e = 0.8$). For the pure water droplet (Figure~\ref{fig:fig4}a), the contact line depins after $t/t_e = 0.2$, leading to a reduction in wetting diameter. In contrast, for the nanoparticle-laden droplet (Figure~\ref{fig:fig4}b), the wetting diameter remains constant up to $t/t_e = 0.8$, indicating persistent pinning.

Figure~\ref{fig:fig3}(b) presents the droplet lifetimes for completely pinned and partially pinned configurations, normalized by the corresponding isolated droplet lifetime. The normalized lifetime, denoted by $\tau$, is defined as
\begin{equation} \label{j:eq1}
\tau = \frac{t_e(N)}{t_e(1)},
\end{equation}
where $t_e(N)$ is the lifetime of a configuration containing $N$ droplets in a given evaporation mode, and $t_e(1)$ is the lifetime of an isolated droplet evaporating in the same mode as $t_e(N)$. The parameter $\tau$ therefore quantifies the relative increase in droplet lifetime with increasing $N$ for both completely pinned and partially pinned droplets. It can be seen in Figure~\ref{fig:fig3}(b) that the value of $\tau$ is consistently higher for completely pinned cases, regardless of the number of droplets in the configuration, indicating that complete pinning enhances the shielding effect more strongly than in partially pinned cases. For a five-droplet arrangement, the lifetime is $1.89$ times that of the isolated droplet in the partially pinned mode and $2.24$ times in the completely pinned mode. This difference arises because the shielding effect depends not only on the number of droplets but also on their proximity and the contact line dynamics exhibited during evaporation. Since the contact radius of a completely pinned droplet remains constant throughout evaporation, all droplets in the configuration maintain the same proximity to one another, unlike in the partially pinned case. This results in a prolonged and more intense shielding effect for completely pinned droplets. In contrast, for partially pinned droplets, the receding contact line reduces droplet proximity over time, thereby weakening the shielding effect as evaporation progresses. Thus, during evaporation, the extent of shielding is governed by the interplay between the number of droplets in the configuration and the evaporation mode (completely pinned or partially pinned droplets).

It should be noted that the evaporation rate of an isolated partially pinned droplet is inherently lower than that of an isolated completely pinned droplet. This is because completely pinned droplets retain a larger perimeter and free surface area even in the later stages of evaporation, whereas both parameters decrease over time for partially pinned droplets, slowing their evaporation. However, as the number of droplets increases, the enhanced shielding in completely pinned configurations counteracts their intrinsically higher evaporation rates. Consequently, the lifetime difference between completely pinned and partially pinned droplets decreases with increasing droplet number. This effect is illustrated in Figure~\ref{fig:fig3}(c), which presents the relative variation in lifetime between partially pinned and completely pinned droplets for each configuration, denoted as $t_{\mathrm{dev}}$ (\%) and defined as
\begin{equation} \label{j:eq2}
t_{dev}(\%) = \frac{t_{e, ~partially-pinned}-{{t_{e, ~completely-pinned}}}}{t_{e, ~partially-pinned }}\times100
\end{equation}
where $t_{e, ~partially-pinned}$ and $t_{e,~completely-pinned}$ denote the lifetimes of partially pinned and completely pinned droplets, respectively, for a given configuration.

For an isolated droplet, the lifetime in the completely pinned mode is about 21\% shorter than in the partially pinned mode. This difference decreases with increasing droplet number, reaching approximately 6.67\% for the five-droplet configuration. This trend indicates that the number of droplets and the evaporation mode (completely pinned or partially pinned) act in competition to determine the overall evaporation rate of the configuration. \ks{In completely pinned droplets, as evaporation progresses, the vapor concentration progressively increases near the droplet surface.} In contrast, in partially pinned droplets, vapor is more evenly distributed in the space between droplets, away from their surfaces. Consequently, completely pinned droplets are exposed to higher local vapor concentrations due to their fixed contact lines, whereas the receding contact lines of partially pinned droplets reduce this effect over time. The combined influence of droplet number and evaporation mode alters the shielding effect significantly, thereby producing the observed differences in evaporation rates. Next, we examine the top-view droplet profiles during evaporation for the different configurations.

\subsection{Top-view Profiles}

\begin{figure}
\centering
\vspace{0 mm}\includegraphics[width=0.9
\textwidth]{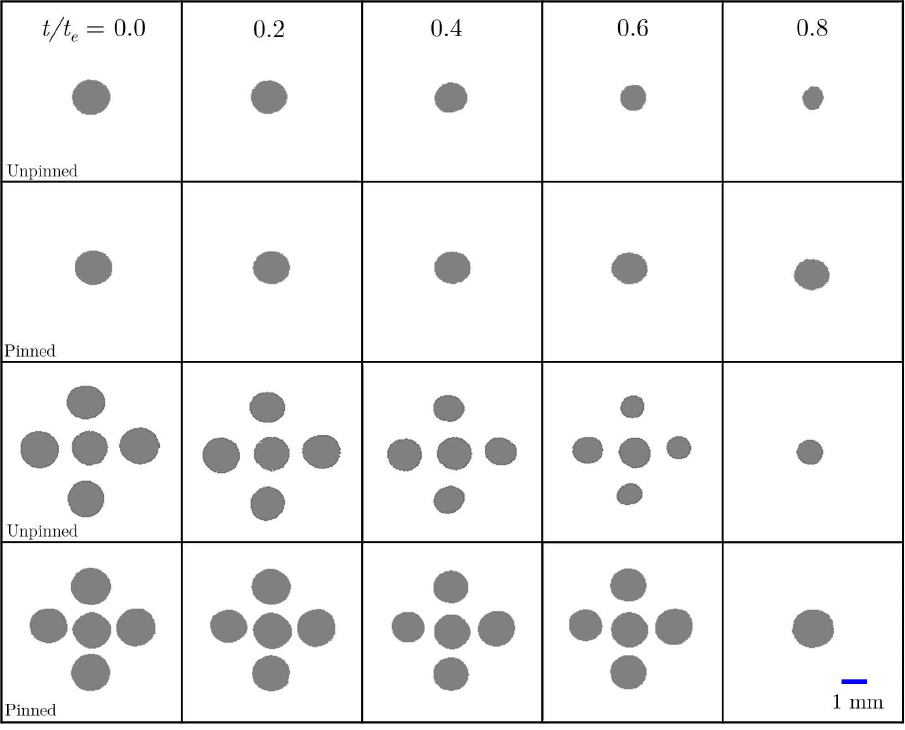} \hspace{2mm}\\
\caption{Temporal evolution of top-view droplet profiles, with the first and third rows corresponding to partially pinned droplets, and the second and fourth rows depicting completely pinned droplets. A scale bar is provided in the bottom-right panel.} 
\label{fig:fig5}
\end{figure}

In this section, we examine the top view of droplets arranged in different configurations. Figure~\ref{fig:fig5} illustrates the temporal evolution of the top-view profiles for both partially pinned and completely pinned cases at different normalized evaporation times.

For an isolated partially pinned droplet, the diameter remains nearly constant until $t/t_e = 0.2$, after which the contact line begins to depin. With further evaporation, the wetting diameter decreases, leading to partially pinned behavior. The five-droplet arrangement exhibits similar behavior, except for the central droplet, which maintains a slightly larger spread than the side droplets for most of the evaporation time. This is attributed to the shielding effect of the surrounding side droplets, which increases the vapor concentration in the azimuthal direction around the central droplet. For the side droplets, the vapor concentration is higher on the surface facing the central droplet than on the opposite side, resulting in a lower and slightly non-uniform azimuthal vapor concentration overall. This difference increases the evaporation rate of the side droplets, causing them to evaporate faster than the central droplet. In the completely pinned cases, the top-view spread remains constant for all droplets in a given arrangement. The droplets remain in close proximity throughout the evaporation process, prolonging the shielding effect and intensifying it for most of the evaporation time. Then, we analyse the evaporation dynamics by examining the temporal evolution of the height, wetting diameter, contact angle, and volume of droplets in different configurations during evaporation, as obtained from side-view imaging using a CMOS camera. These results serve as the basis for developing the theoretical model discussed in the subsequent section.

\subsection{Side-view Profiles: Evaporation dynamics}

We examine the temporal evolution of the geometrical parameters for both completely pinned and partially pinned droplets, namely the normalized height ($h/h_0$), normalized wetted diameter ($D/D_0$), contact angle ($\theta$ in degrees), and normalized volume ($V/V_0$) of droplets evaporating in different configurations. The initial height, wetted diameter, contact angle, and volume of the droplet are denoted by $h_0$, $D_0$, $\theta_0$, and $V_0$, respectively. The procedure for extracting contour profiles from CMOS camera data has been described in the experimental section. Due to imaging limitations, accurate estimation of these parameters is possible only up to $t/t_e = 0.8$. The variations are presented for isolated droplets as well as for two-, three-, and five-droplet configurations. Across all cases, the initial values of $h_0$, $D_0$, $\theta_0$, and $V_0$ are $1.00 \pm 0.02$ mm, $1.60 \pm 0.05$ mm, $100.2 \pm 2.5^\circ$, and $1.40 \pm 0.10$ μl, respectively.

\begin{figure}[h]
\centering
\hspace{0.5cm}{\large (a)} \hspace{7.1cm}{\large (b)} \\
 \includegraphics[width=0.45\textwidth]{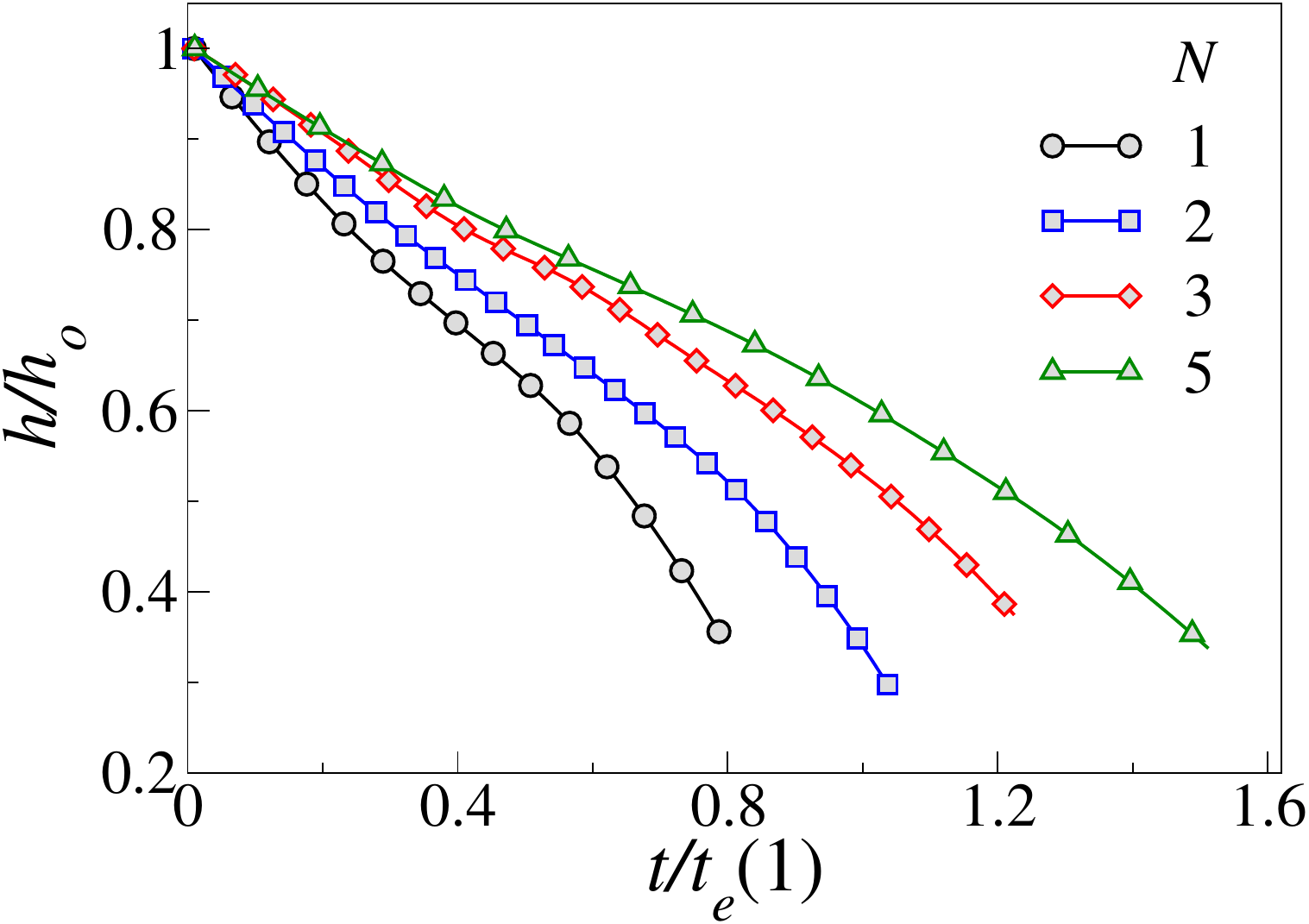} \hspace{2mm} \includegraphics[width=0.45\textwidth]{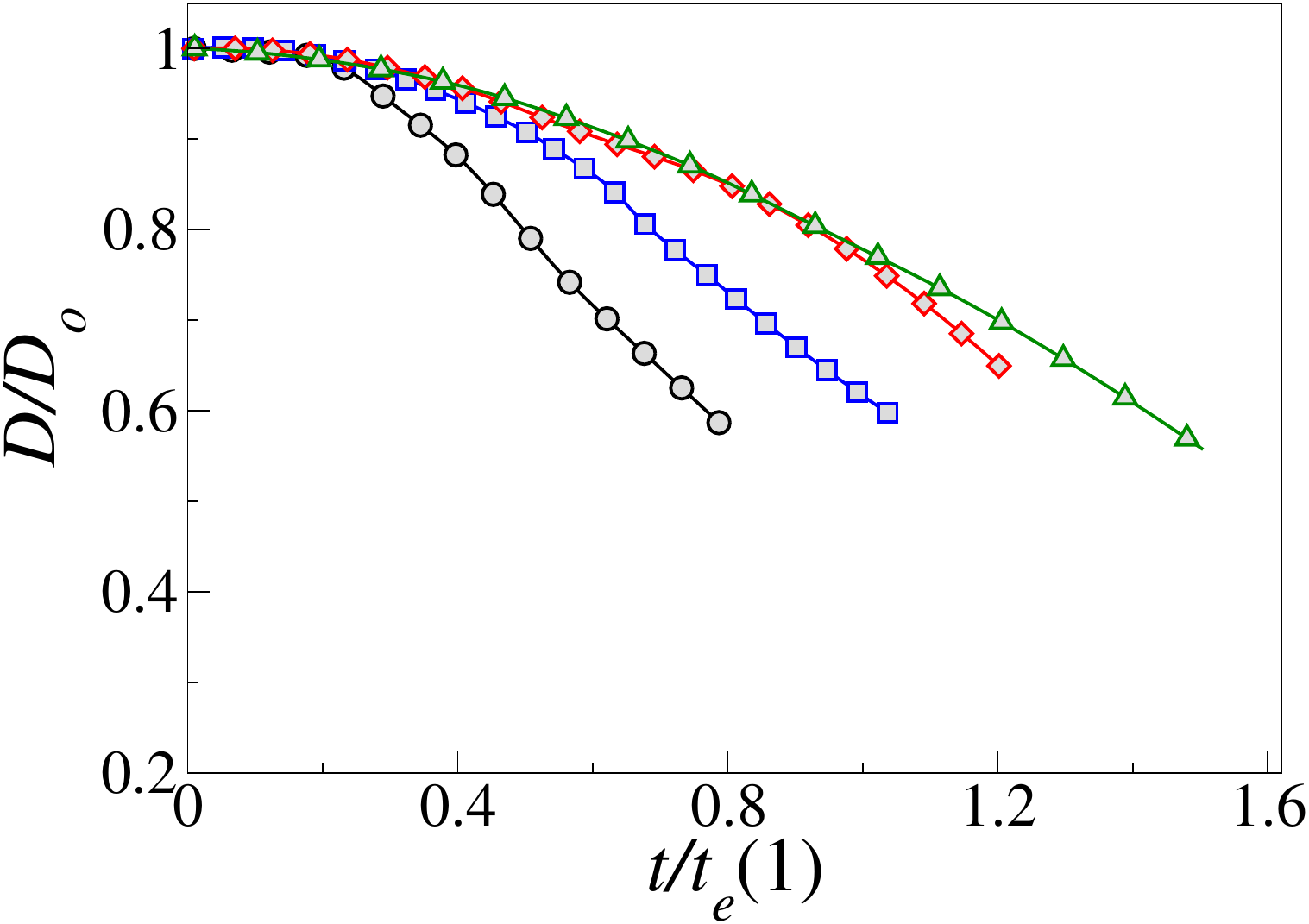}\\
 \hspace{0.5cm}{\large (c)} \hspace{7.1cm}{\large (d)} \\
 \includegraphics[width=0.45\textwidth]{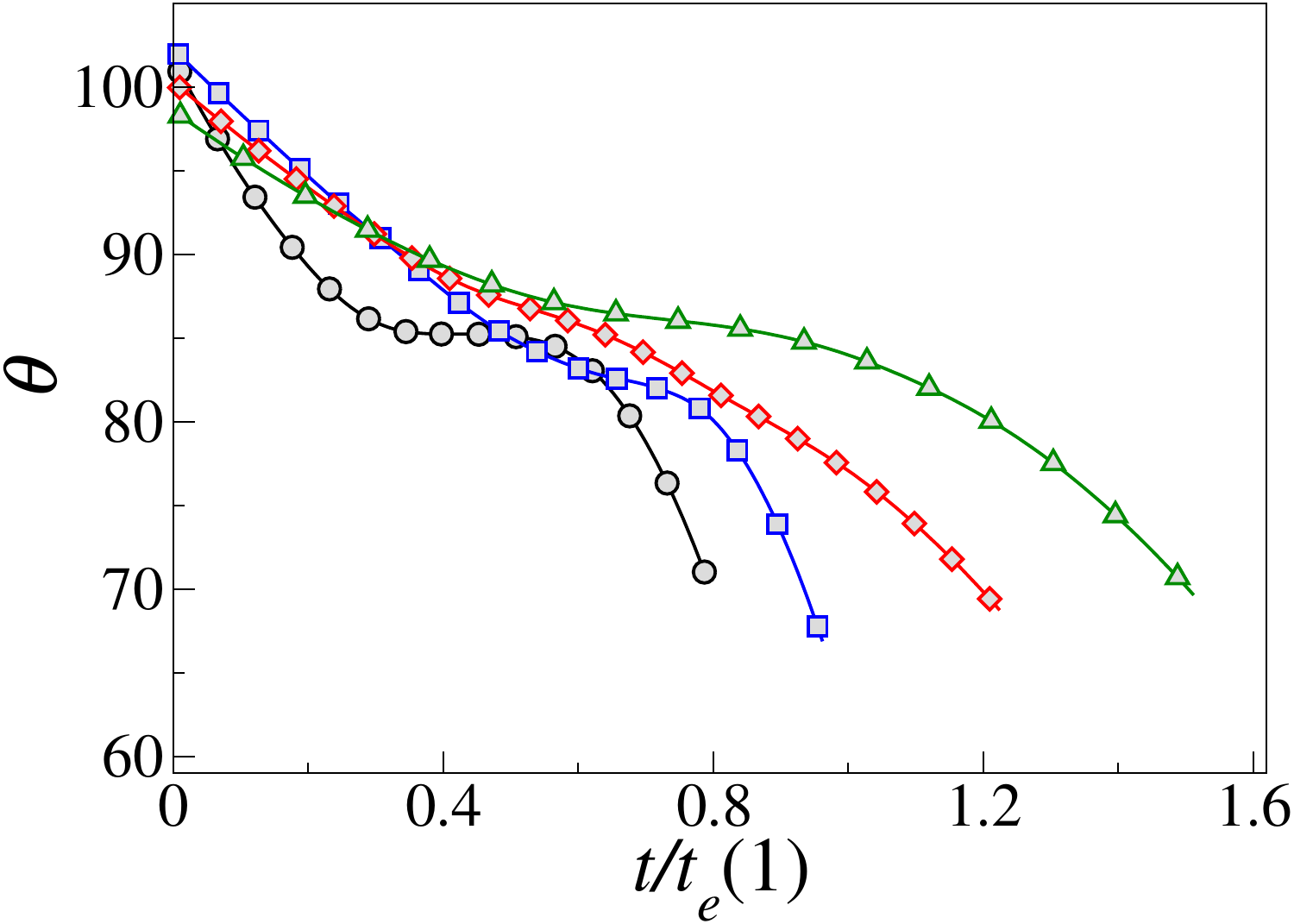} \hspace{2mm} \includegraphics[width=0.45\textwidth]{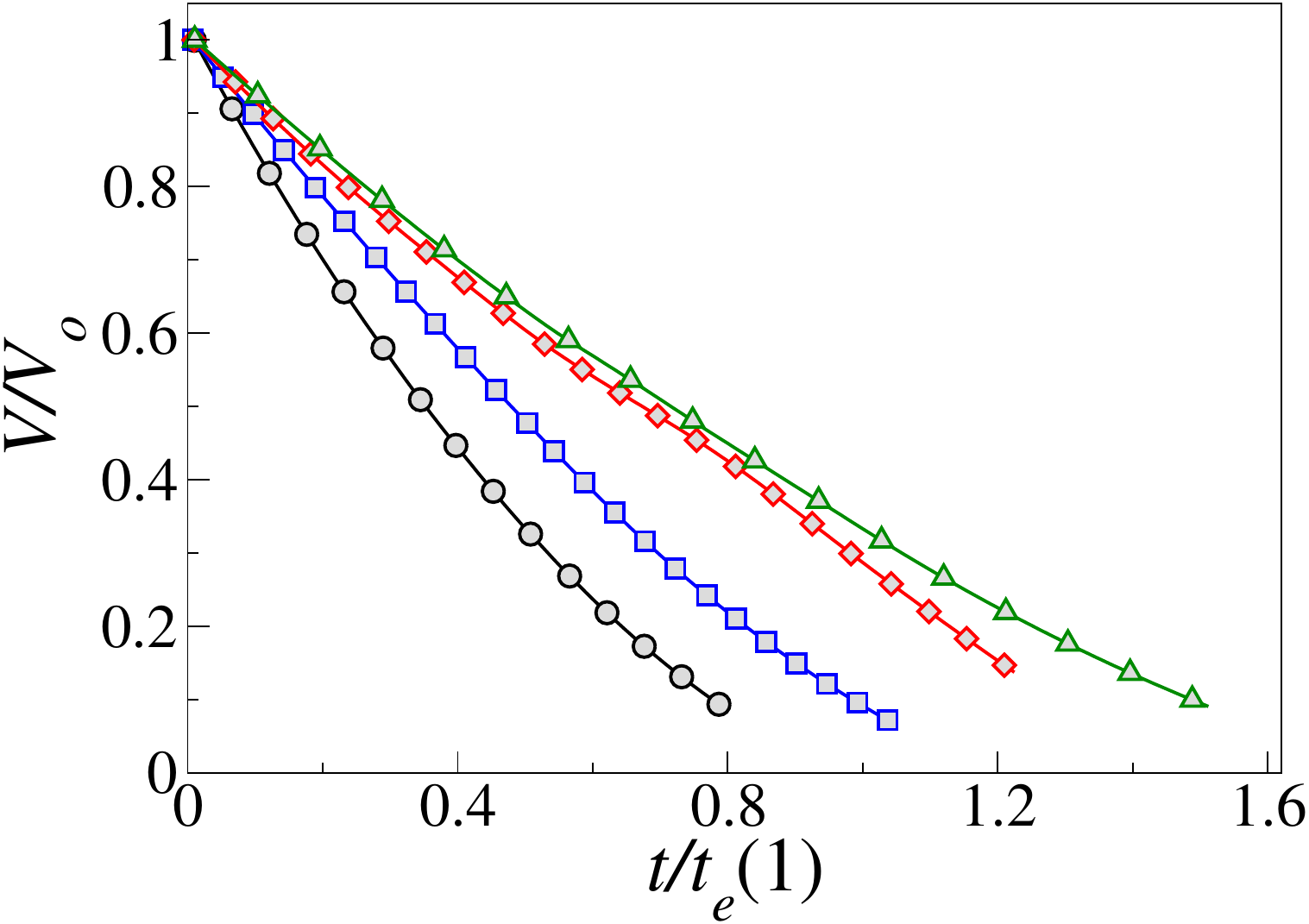}\\
\caption{Variations of (a) normalized height ($h/h_0$), (b) normalized wetted diameter ($D/D_0$), (c) contact angle ($\theta$ in degrees), and (d) normalized volume ($V/V_0$) of partially pinned multiple droplets arranged in different configurations with normalized time $t/t_e(1)$, where $t_e(1)$ denotes the total evaporation time of the corresponding isolated droplet.}
\label{fig:fig6}
\end{figure}

\begin{figure}[h]
\centering
\hspace{0.5cm}{\large (a)} \hspace{7.1cm}{\large (b)} \\
 \includegraphics[width=0.45\textwidth]{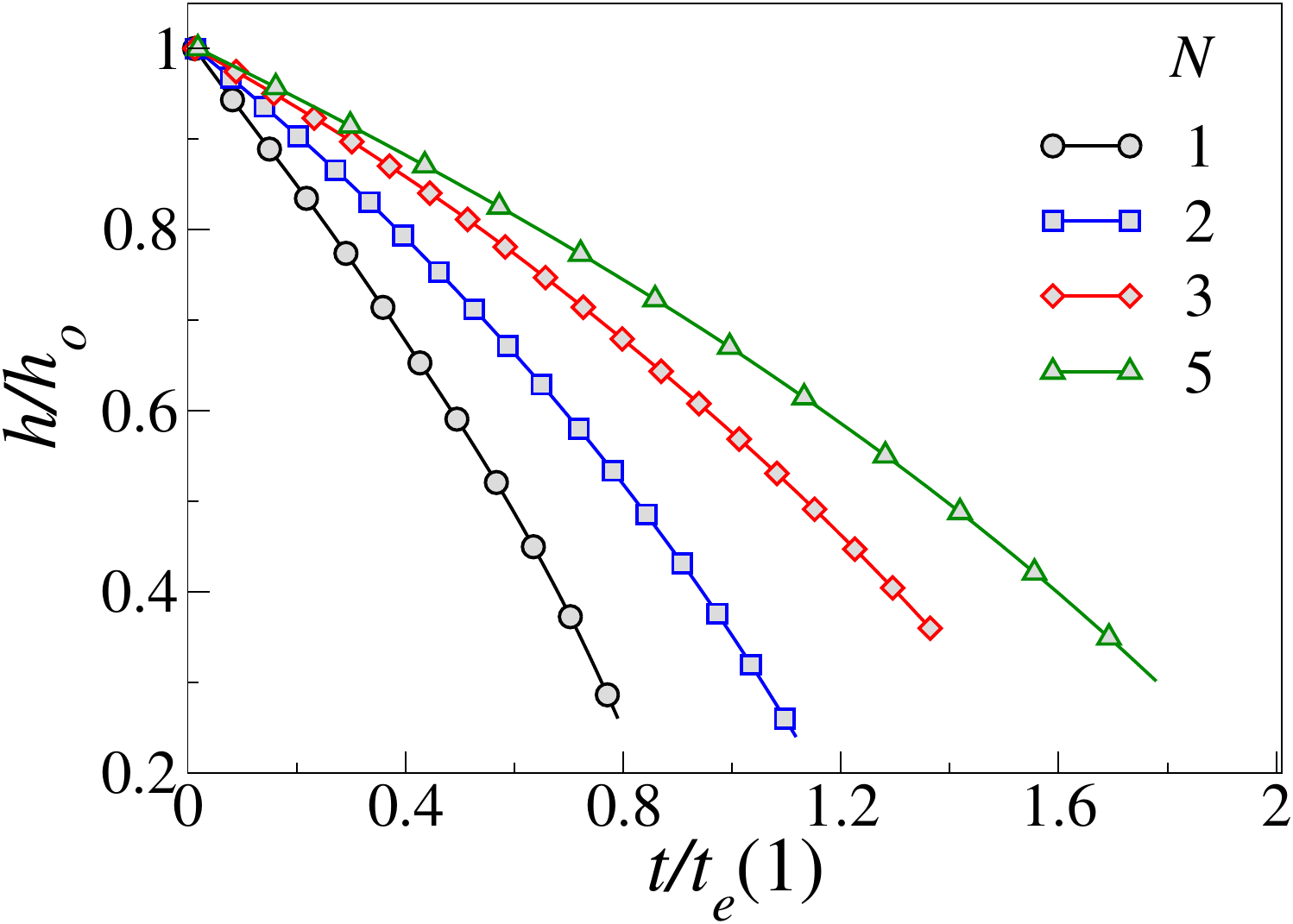} \hspace{2mm} \includegraphics[width=0.45\textwidth]{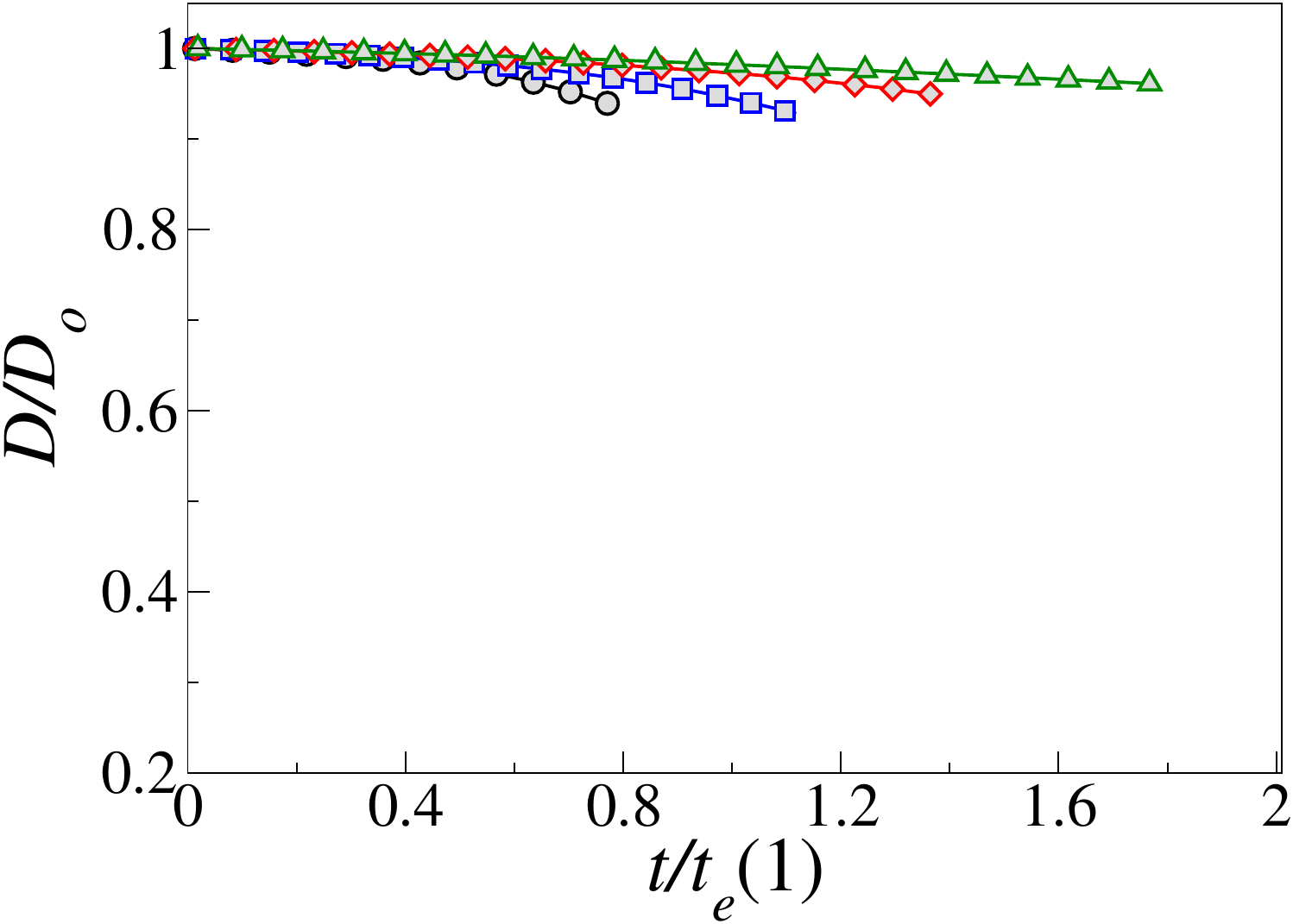}\\
 \hspace{0.5cm}{\large (c)} \hspace{7.1cm}{\large (d)} \\
 \includegraphics[width=0.45\textwidth]{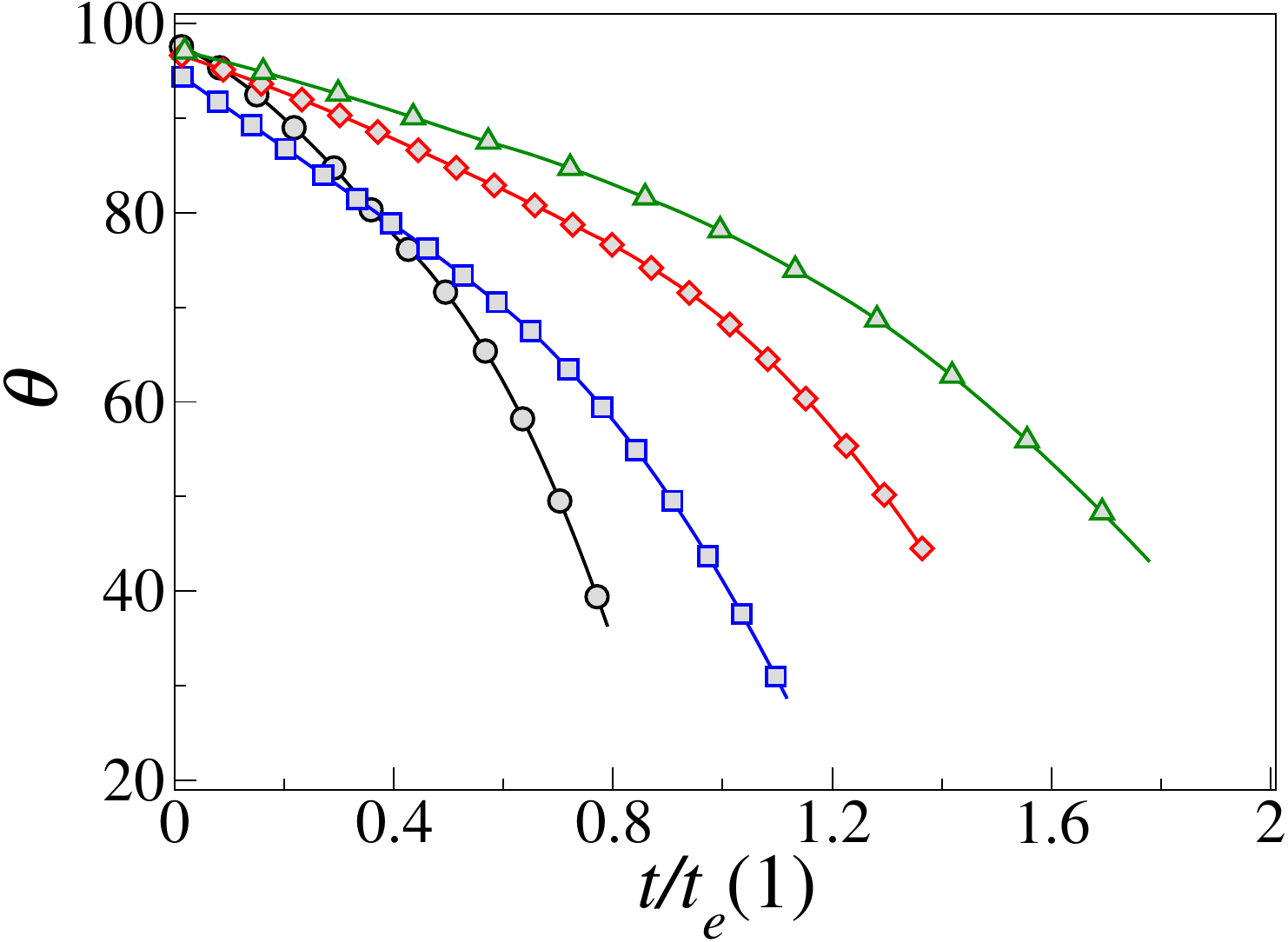} \hspace{2mm} \includegraphics[width=0.45\textwidth]{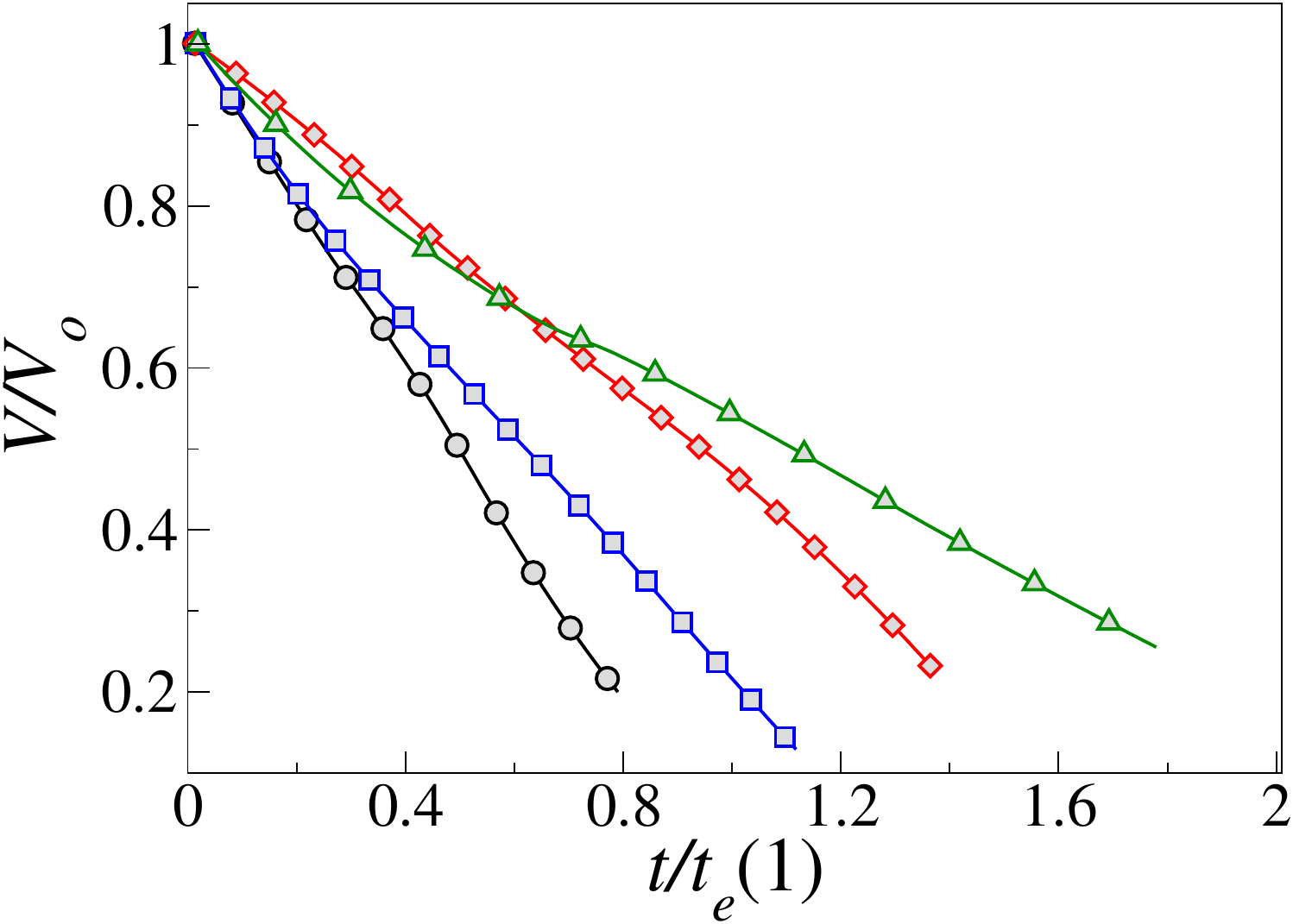}\\
\caption{Variations of (a) normalized height ($h/h_0$), (b) normalized wetted diameter ($D/D_0$), (c) contact angle ($\theta$ in degrees), and (d) normalized volume ($V/V_0$) of completely pinned multiple droplets arranged in different configurations with normalized time $t/t_e(1)$, where $t_e(1)$ denotes the total evaporation time of the corresponding isolated droplet.}
\label{fig:fig7}
\end{figure}

The variation of the droplet geometrical parameters with the evaporation time, normalized by the lifetime of an isolated droplet ($t_{e}(1)$), for the partially pinned cases is shown in Figure~\ref{fig:fig6}. Irrespective of the number of droplets in the configuration, both the height and the normalized volume decrease monotonically with time. The evolution of the wetted diameter indicates that the droplet undergoes two distinct evaporation modes, namely, a constant contact radius (CCR) mode, followed by a monotonic decrease in diameter, characteristic of the constant contact angle (CCA) mode. In the final stage, a mixed mode of evaporation is observed, in which both the wetted diameter and the contact angle decrease simultaneously. Although the evaporation of pure droplets may involve multiple transition modes, we refer to it as the partially pinned mode in this study to clearly distinguish it from the completely pinned mode observed in nanoparticle-laden droplets. The variation of droplet geometrical parameters for completely pinned cases is shown in Figure~\ref{fig:fig7}. Here, due to strong pinning induced by nanoparticles, the height, contact angle, and normalized volume decrease monotonically, while the wetted diameter remains nearly constant throughout most of the evaporation process. The only notable deviation is in the two-droplet configuration, where the wetted diameter varies by about 7\% from its initial value. Although the normalized lifetime of partially pinned droplets is lower than that of completely pinned droplets across all configurations (Figures~\ref{fig:fig6} and \ref{fig:fig7}), the absolute lifetimes of completely pinned droplets are shorter than those of partially pinned droplets. Note that the droplet geometrical parameters are measured over three independent repetitions for a given arrangement, and the averaged values are presented in Figures \ref{fig:fig6} and \ref{fig:fig7}.

\begin{figure}[h]
\centering
\hspace{0.5cm}{\large (a)} \hspace{7.1cm}{\large (b)} \\
 \includegraphics[width=0.45\textwidth]{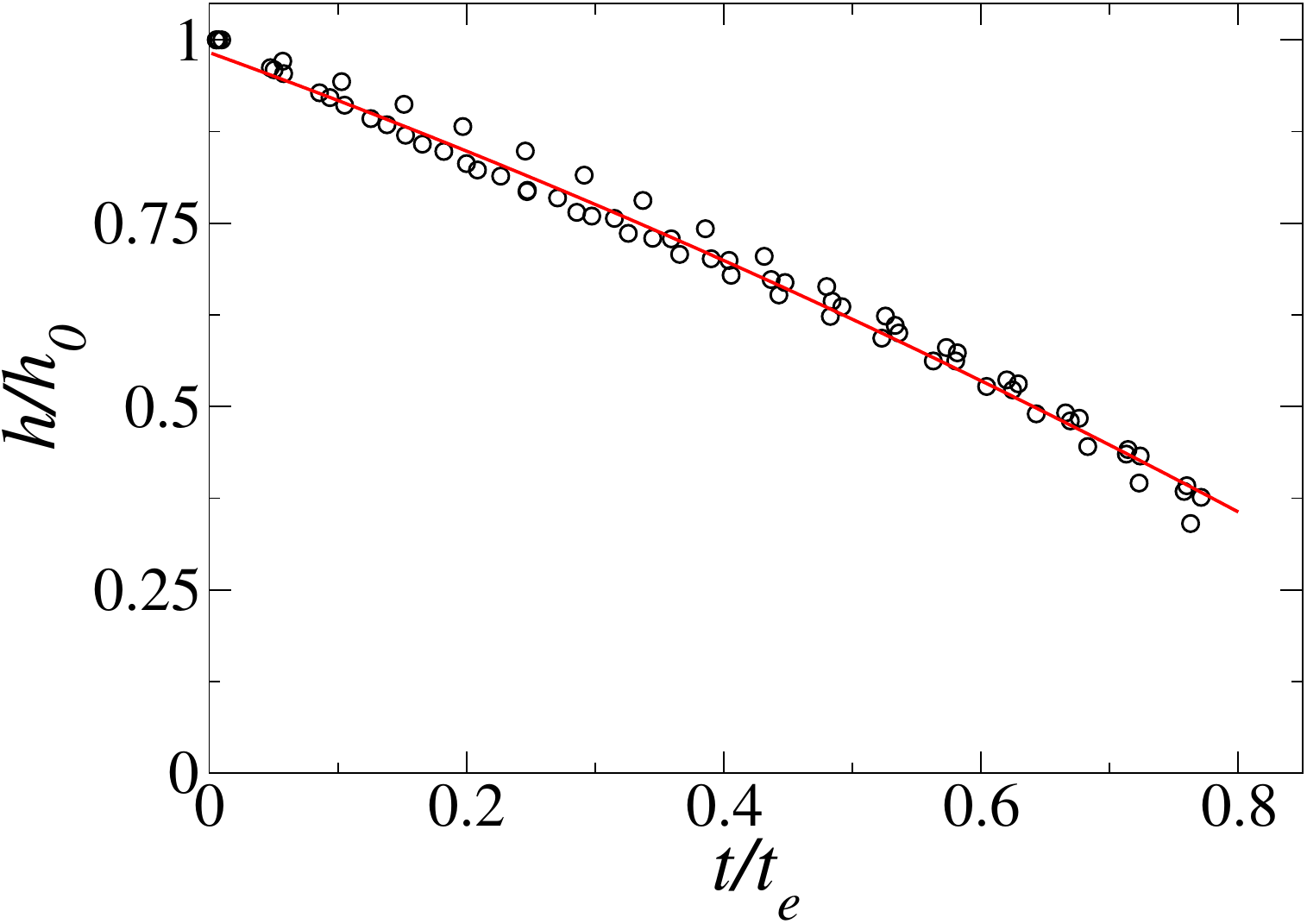} \hspace{2mm} \includegraphics[width=0.45\textwidth]{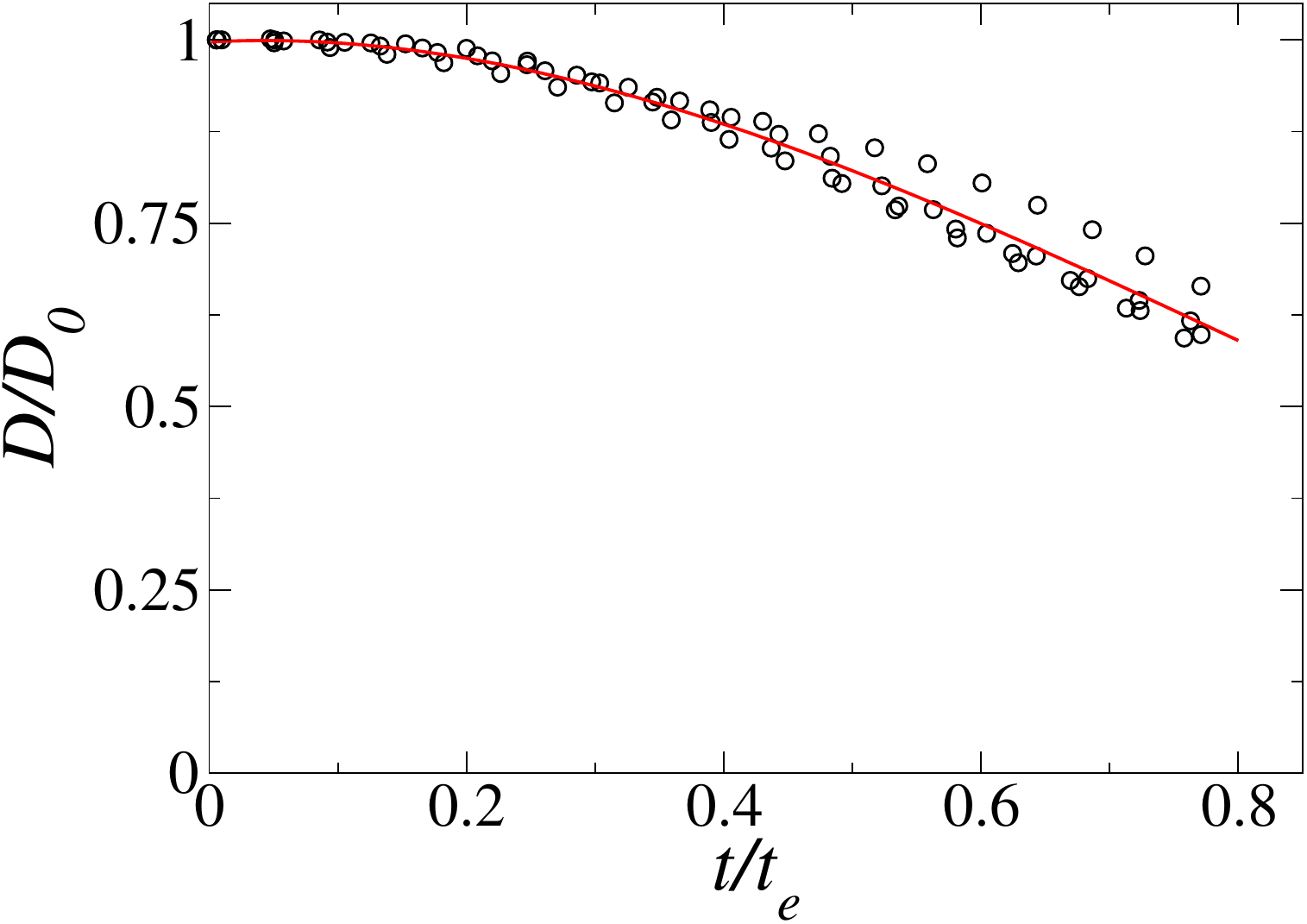}\\
 \hspace{0.5cm}{\large (c)} \hspace{7.1cm}{\large (d)} \\
 \includegraphics[width=0.45\textwidth]{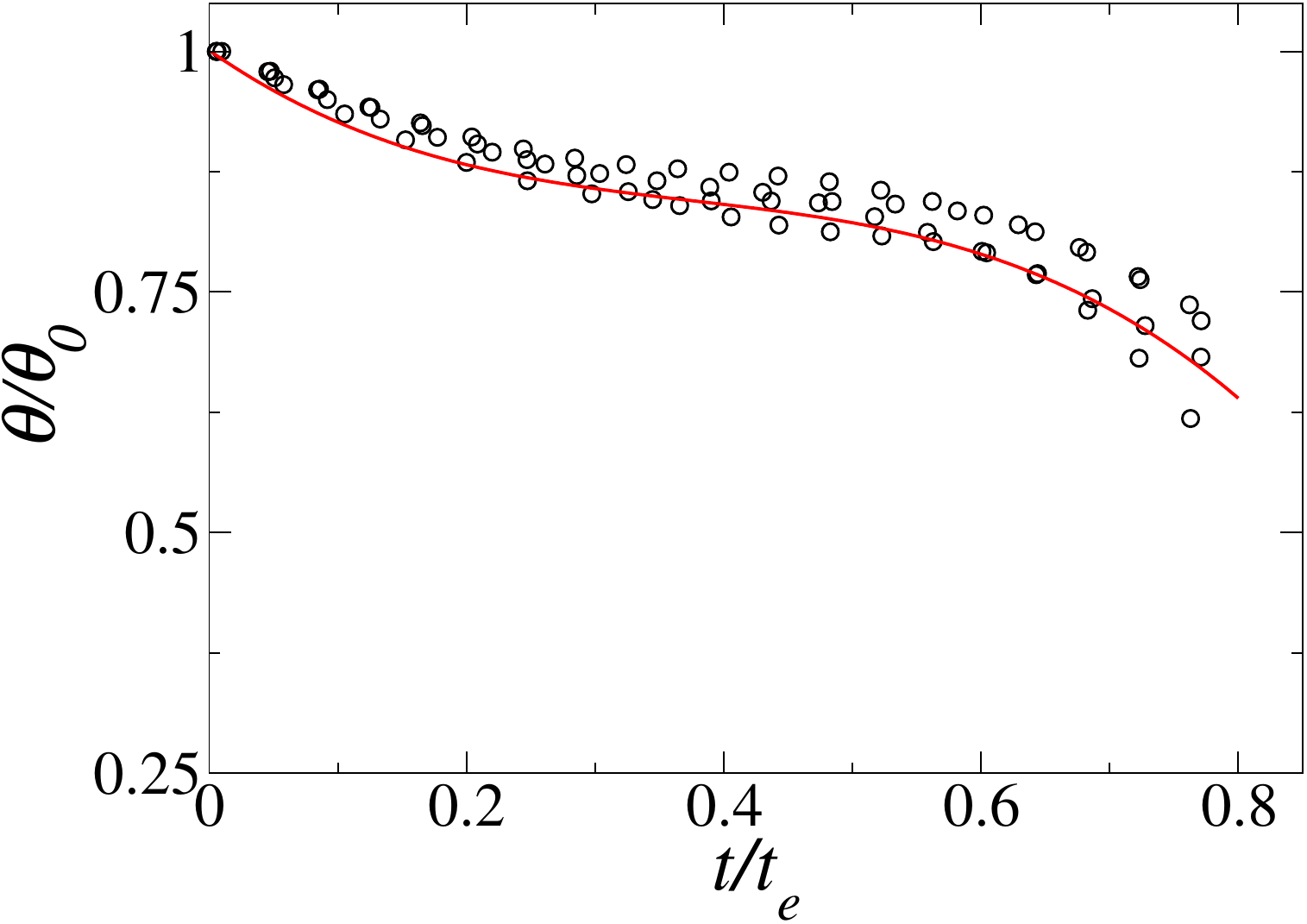} \hspace{2mm} \includegraphics[width=0.45\textwidth]{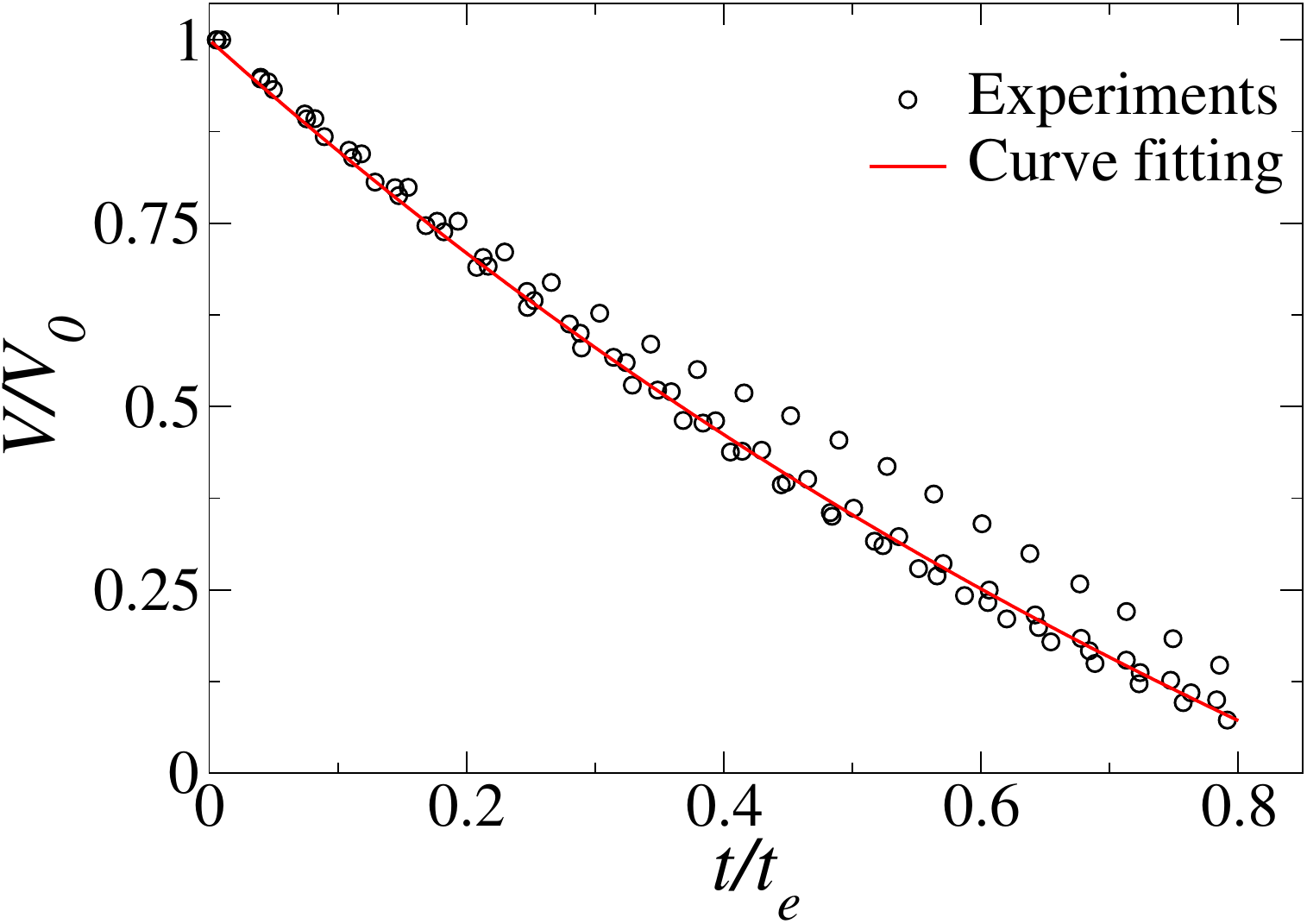}\\
\caption{Variations of (a) normalized height ($h/h_0$), (b) normalised wetted diameter ($D/D_0$), (c) normalized contact angle ($\theta/\theta_0$), and (d) normalized volume ($V/V_0$) of partially pinned multiple droplets arranged in different configurations, plotted against the normalized time with respect to their respective droplet lifetimes ($t/t_e$).}
\label{fig:fig8}
\end{figure}

\begin{figure}[h]
\centering
\hspace{0.5cm}{\large (a)} \hspace{7.1cm}{\large (b)} \\
 \includegraphics[width=0.45\textwidth]{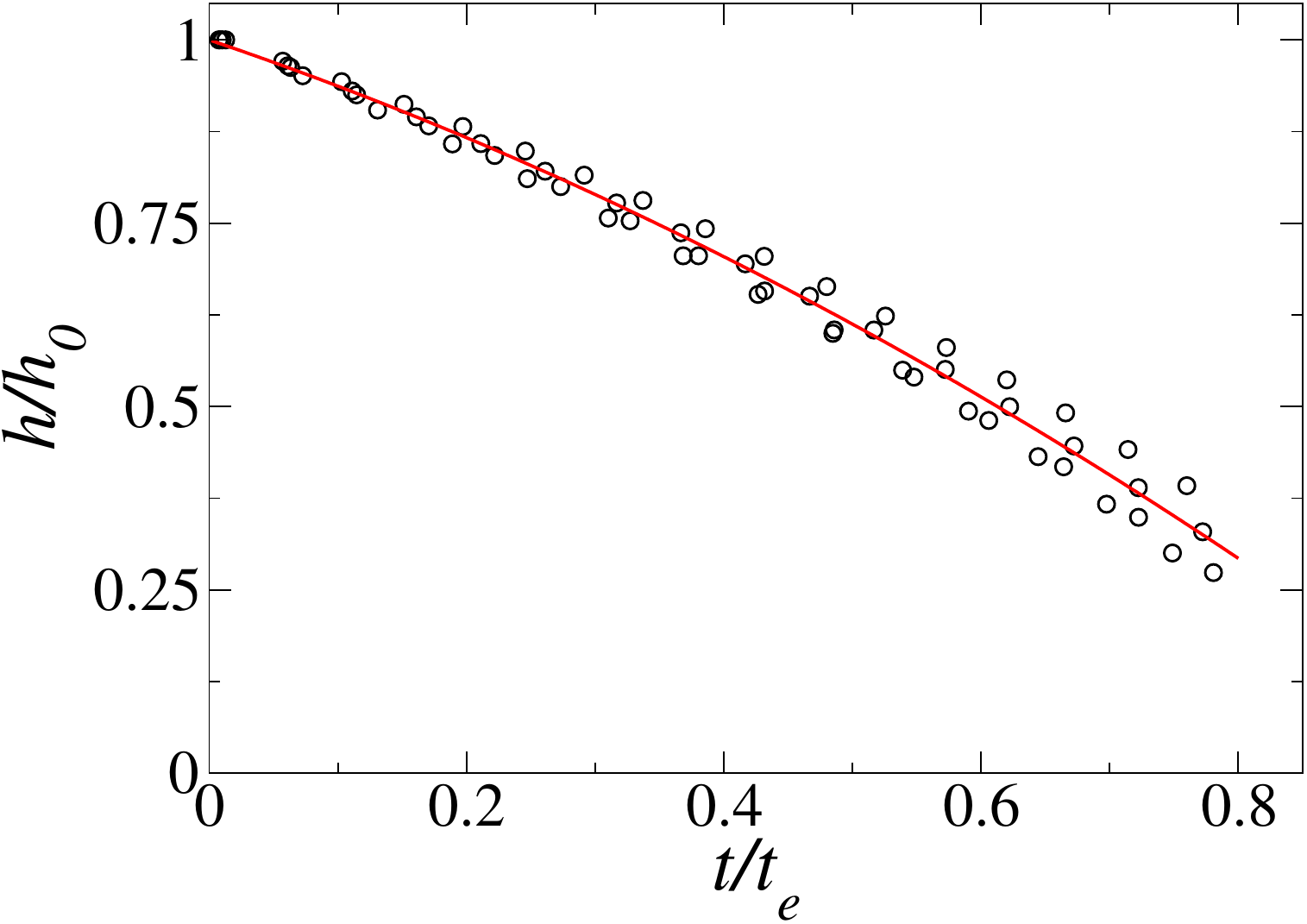} \hspace{2mm} \includegraphics[width=0.45\textwidth]{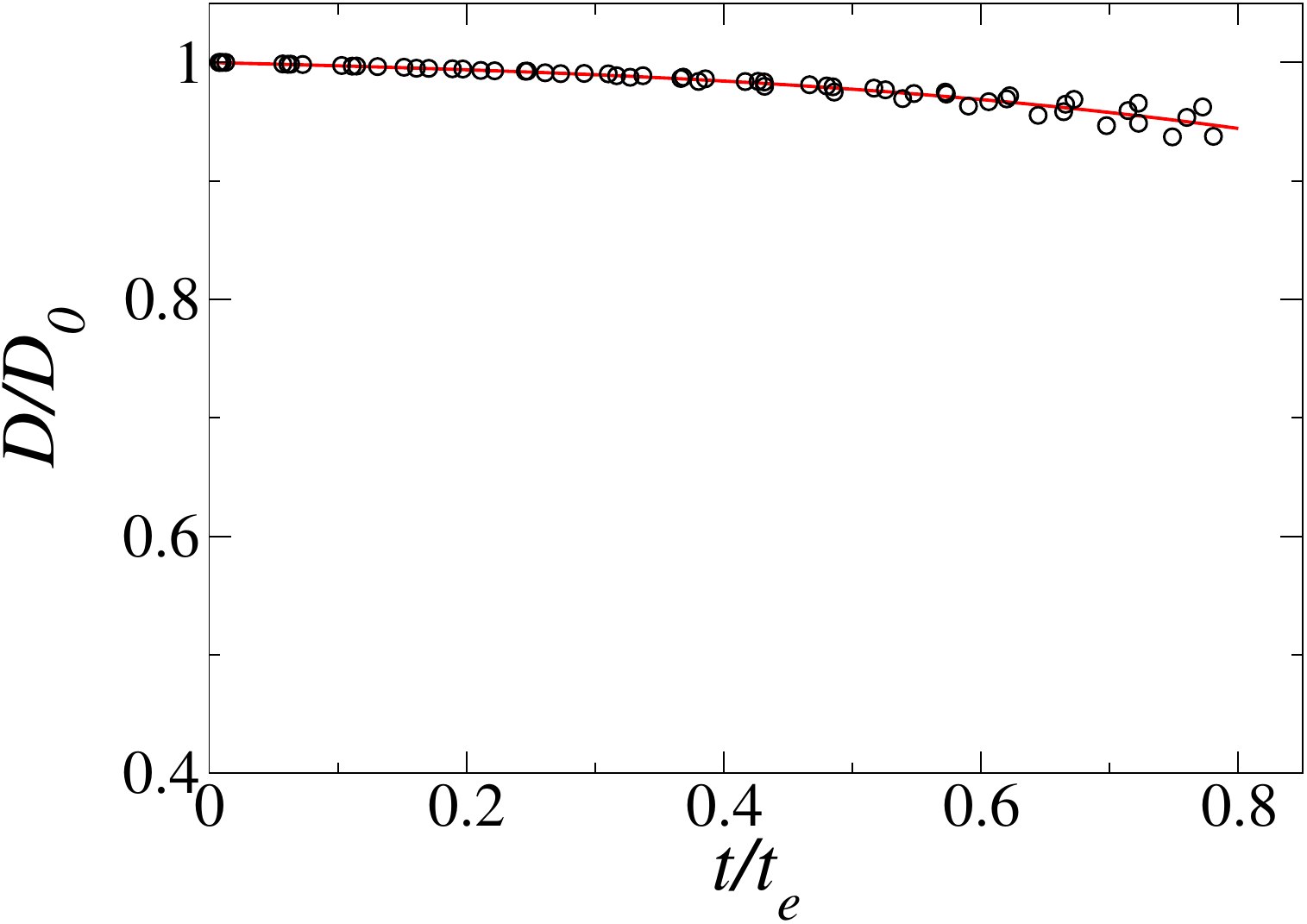}\\
 \hspace{0.5cm}{\large (c)} \hspace{7.1cm}{\large (d)} \\
 \includegraphics[width=0.45\textwidth]{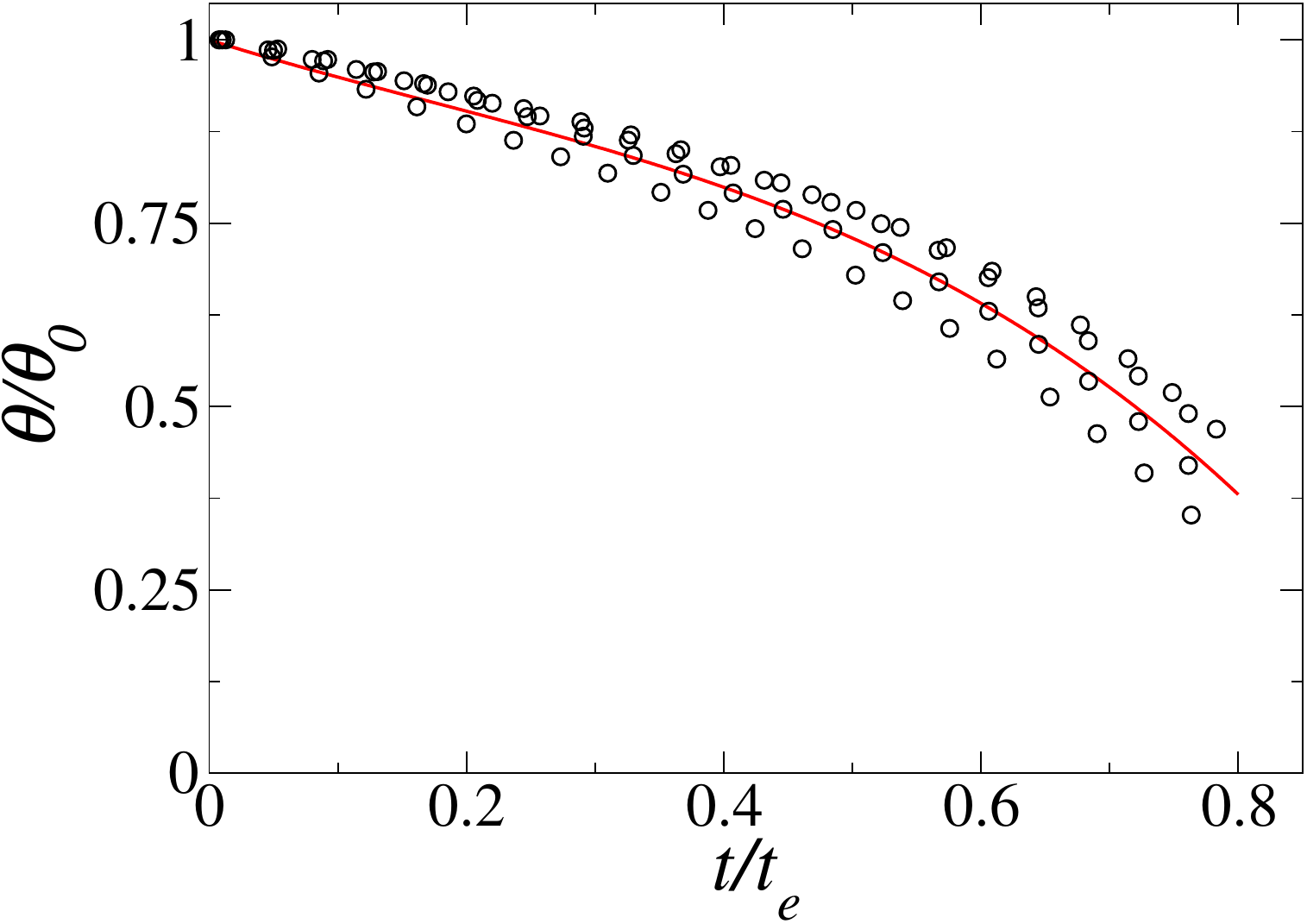} \hspace{2mm} \includegraphics[width=0.45\textwidth]{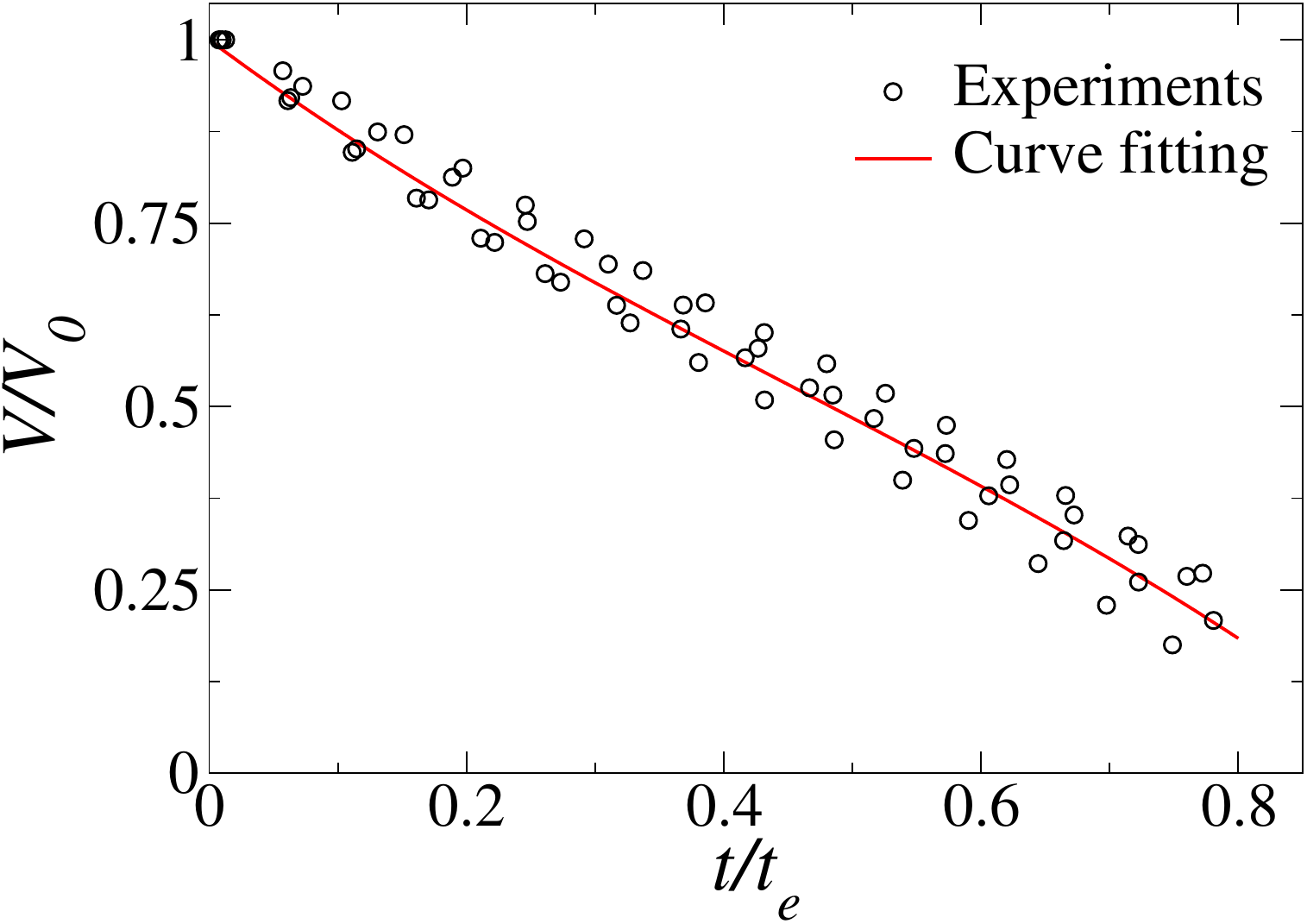}\\
\caption{Variations of (a) normalized height ($h/h_0$), (b) normalized wetted diameter ($D/D_0$), (c) normalized contact angle ($\theta/\theta_0$), and (d) normalized volume ($V/V_0$) of completely pinned multiple droplets arranged in different configurations, plotted against the normalized time with respect to their respective droplet lifetimes ($t/t_e$).}
\label{fig:fig9}
\end{figure}

\begin{figure}
\centering
\vspace{0 mm}\includegraphics[width=0.9
\textwidth]{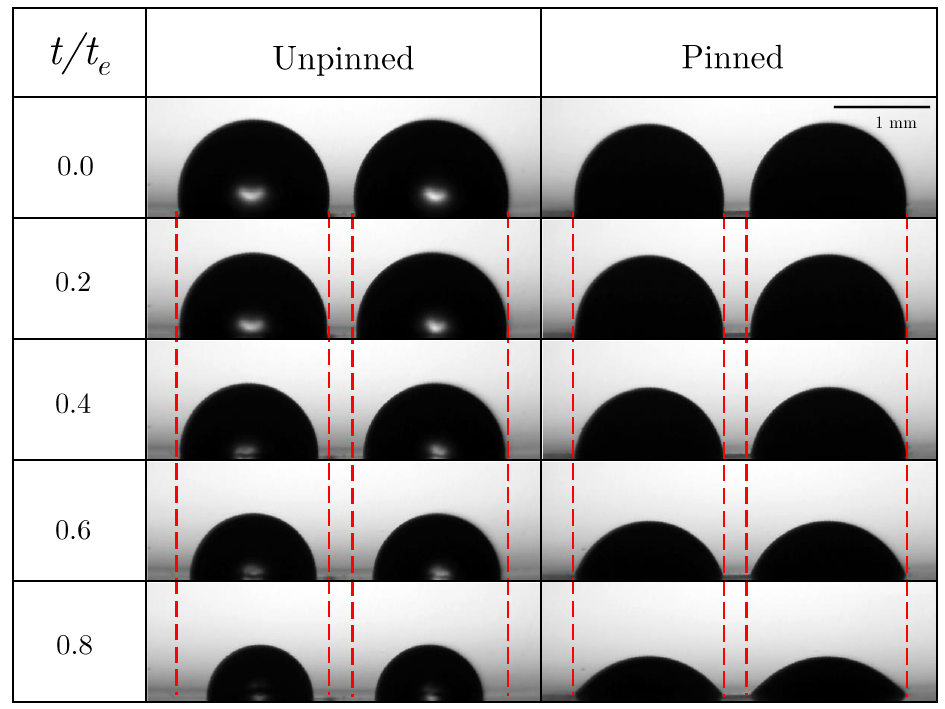} \hspace{2mm}\\
\caption{Temporal evolution of shape of the droplets in the two-droplet configuration for partially pinned and completely pinned contact line conditions.} 
\label{fig:fig10}
\end{figure}

A generalized behavior in droplet parameters is evident for both completely pinned and partially pinned contact line cases when evaporation time is normalized by the respective droplet lifetime ($t_e$). This finding is consistent with earlier studies \cite{dash2014droplet,hatte2019universal,Hari2024}. These trends are presented in Figures~\ref{fig:fig8} and \ref{fig:fig9}, where data points include measurements from both central and side droplets. To capture the average behavior, a polynomial fit (red line) was applied to the experimental data (black circles), with the corresponding fit coefficients and $R^2$ values provided in Supplementary Tables S3 and S4 for the partially pinned and completely pinned droplets, respectively. The close agreement between the experimental data and the fitted curves confirms that the contact line dynamics exhibit similar scaling behavior across different droplet arrangements when normalized by their respective lifetimes. This pinning and depinning behavior of the contact line is illustrated in Figure \ref{fig:fig10} for a two-droplet configuration. For the partially pinned case, up to $t/t_e = 0.2$, the droplets maintain a constant wetted diameter, after which the contact line begins to depin. During this depinning stage, the contact angle remains nearly constant until $t/t_e = 0.8$. In contrast, the completely pinned case maintains a constant wetted diameter throughout the entire evaporation process, accompanied by a continuous decrease in the contact angle. To summarise, Figures~\ref{fig:fig8} and \ref{fig:fig9} show that, irrespective of whether the droplet is central or side, the normalized evolution of droplet parameters with respect to its own lifetime exhibits a generalized behavior. This indicates that, within a given configuration, contact line dynamics remain unaffected by the droplet’s relative position in the array. The only distinction lies in the evaporation rate, which is lower for the central droplet than that of the side droplets.

\subsection{Theoretical Modeling}

A theoretical model, formulated based on the previous study \cite{Hari2024}, is established to predict the experimental results. The effect of gravity on the droplet shape is neglected since the contact radius is smaller than the capillary length scale. The Bond number, $\text{Bo} = \rho g R^{2}/\sigma$, for the droplets considered in this study is approximately 0.14. Hence, the droplets are assumed to retain a spherical-cap profile for volume estimation during evaporation. The vapor concentration at the liquid–air interface is taken as the saturation value, $C_{s}(T_{d})$, where $T_{d}$ is the surface-averaged droplet temperature accounting for evaporative cooling. The ambient vapor concentration far from the droplet is $H C_{s}(T_{\infty})$, where $H$ is the ambient relative humidity and $T_{\infty}$ is the ambient temperature. The evaporative cooling effect of the droplet is quantified using infrared imaging, which provides the reduced droplet temperature. First, the temperature profile of an isolated evaporating droplet is obtained and then averaged over its surface area to estimate a uniform reduced surface temperature, $T_{d}$ \cite{Hari2024}. Although this model is primarily applicable to thin droplets, previous studies have shown satisfactory agreement even for high contact angle scenarios \cite{khilifi2019study}. The instantaneous evaporation rate of an isolated droplet is expressed as \cite{hu2014effect}:
\begin{equation}\label{j:eq3}
\dot{m}_{iso} = \pi RD_{v}M_{w}\left [ C_{s}\left ( T_{d} \right )- HC_{s}\left ( T_{\infty} \right )\right]g\left ( \theta  \right ),
\end{equation}
where 
\begin{equation}\label{j:eq4}
g \left ( \theta  \right ) = \frac{\sin \theta }{1+\cos \theta } + 4\int_{0}^{\infty}\frac{1+\cosh 2\theta \tau }{\sinh 2\pi \tau }\tanh \left [ \left ( \pi -\theta  \right ) \tau \right ]d\tau.
 \end{equation}
Here, $R$ is the contact radius of the droplet, $D_{v}$ is the vapor diffusion coefficient, and $M_{w}$ is the molecular weight of the fluid. The fluid properties are provided in Supplementary Table S5. The evaporation rate of droplets in an array with $N$ droplets can be determined as \cite{rehman2023effect}
\begin{equation} \label{j:eq5}
\dot{m_i} = \dot{m}_{iso} - \frac{2}{\pi }\sum_{\substack{k=1 \\ k \neq i}}^{N}\dot{m}_{k}\arcsin\left (\frac{R_{i}}{D_{i,k}}\right),
\end{equation}
where $\dot{m_i}$ denotes the evaporation rate of the droplet under consideration, $\dot{m}_{iso}$ is the evaporation rate of an isolated droplet, and $\dot{m}{k}$ represents the evaporation rate of the $k^{th}$ surrounding droplet. Here, $R_i$ is the contact radius of the droplet under consideration, and $D_{i,k}$ is the center-to-center distance between the droplet under consideration and the $k^{th}$ surrounding droplet in the array.

To determine the lifetime of the droplets, Eq. (\ref{j:eq5}) is solved simultaneously for a system of $N$ linear equations, following the approach of \citet{edwards2021interferometric}.
\begin{equation} \label{j:eq6}
  \dot{\textbf{m}}_{iso}=\left[\psi\right] \dot{\textbf{m}},
\end{equation}
where $\dot{\textbf{m}}$ is the evaporation rate vector of the droplets in the configuration, $\dot{\textbf{m}_{iso}}$ is the evaporation rate vector of the droplets if they were isolated, and the inverse matrix, $[\psi] ^{-1}$ is known as the suppression matrix of order $N \times N$, where $N$ is the total number of droplets. The elements of $[\psi]$ are given as, 
\begin{equation}
\psi_{i,j}  = 1, ~~ {\rm when} ~~ i = j, \label{j:eq7}
\end{equation}
\begin{equation}
\psi_{i,j}  =  \frac{2}{\pi}\arcsin\left( \frac{R_i}{D_{ij}} \right) ~~ {\rm when} ~~ i \neq j, \label{j:eq8}
\end{equation}
where $i,j = 1,2, \cdot \cdot \cdot N$ are the counters identifying the droplets, $R_i$ is the radius of the $i^{th}$ droplet and $D_{i,j}$ is the center-to-center distance between the $i^{th}$ and the $j^{th}$ droplets. 
The expression for the matrix $[\psi]$ for a five-droplet configuration is provided in the Supplementary Material (see, section Suppression matrix). Using this framework, we theoretically predicted the evaporation rates and lifetimes of the central droplets. It should be noted that the contact radius and contact angle of the central droplet can be reliably estimated only up to $t/t_{e} = 0.8$, owing to the complex morphologies that arise at $t/t_{e} > 0.8$. Therefore, the evaporation time is calculated up to 80\% of the total lifetime.

\begin{figure}[h]
\centering
\hspace{0.5cm}{\large (a)} \hspace{7.1cm}{\large (b)} \\
 \includegraphics[width=0.45\textwidth]{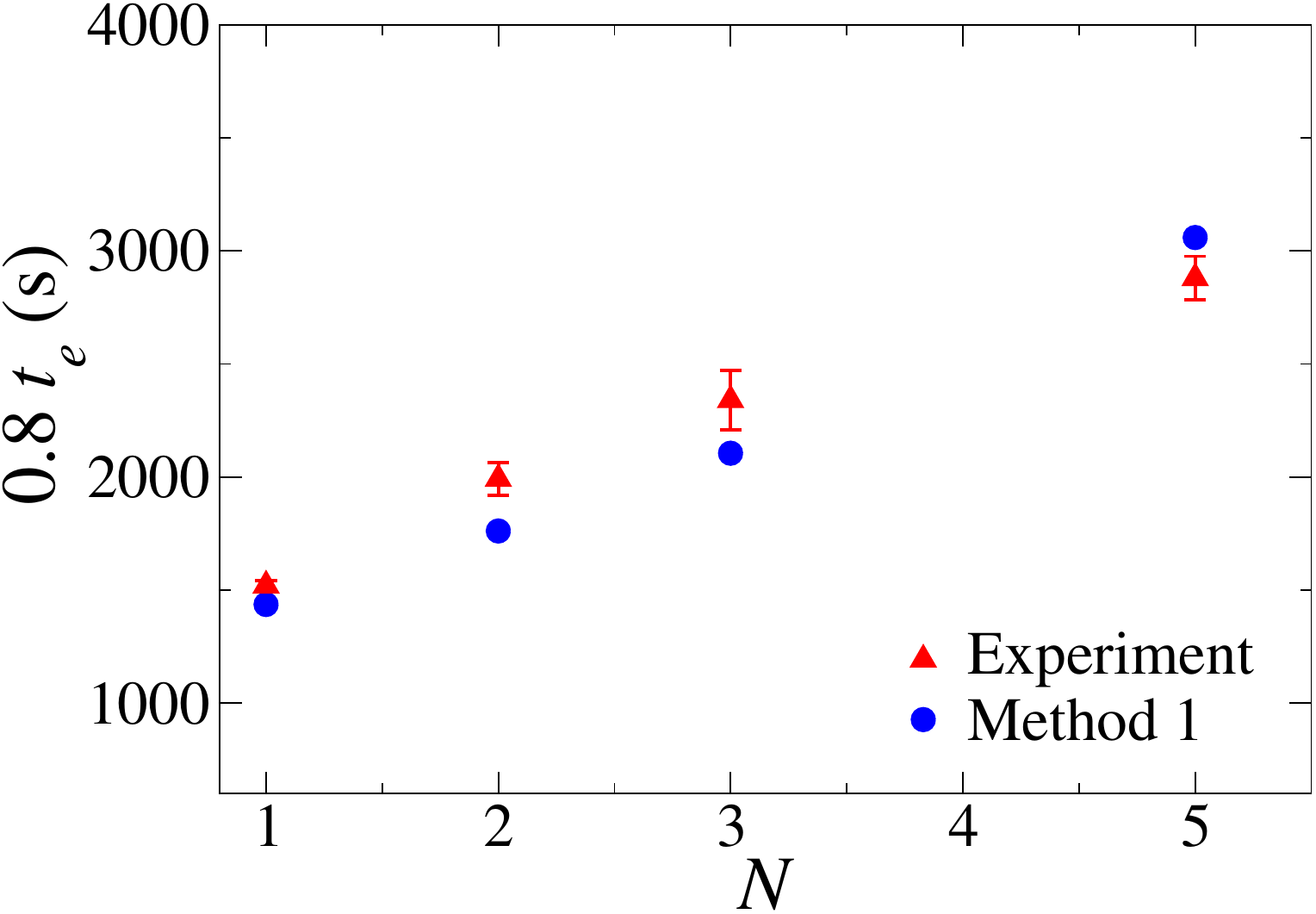} \hspace{2mm} \includegraphics[width=0.45\textwidth]{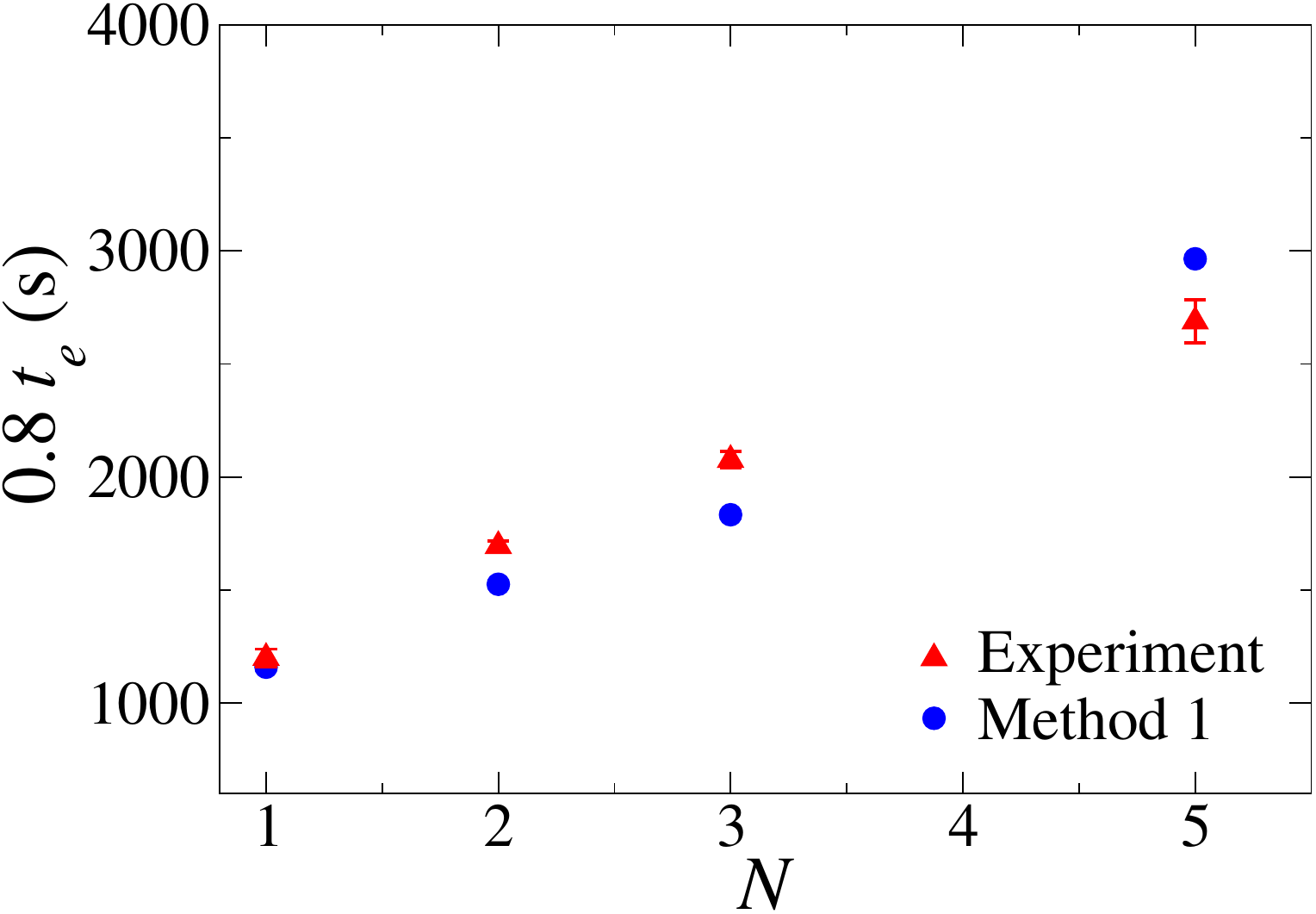}\\
 \hspace{0.5cm}{\large (c)} \hspace{7.1cm}{\large (d)} \\
 \includegraphics[width=0.45\textwidth]{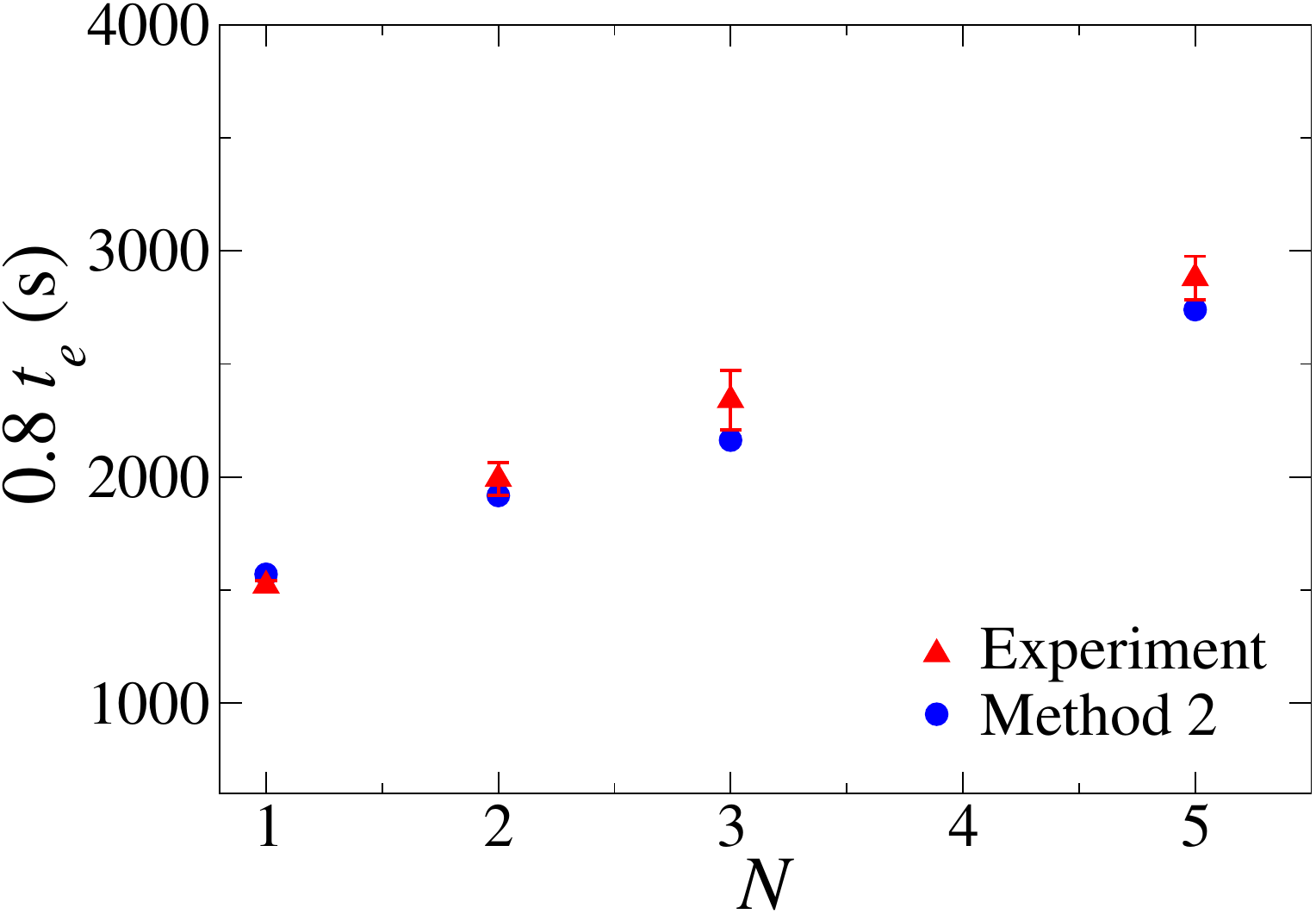} \hspace{2mm} \includegraphics[width=0.45\textwidth]{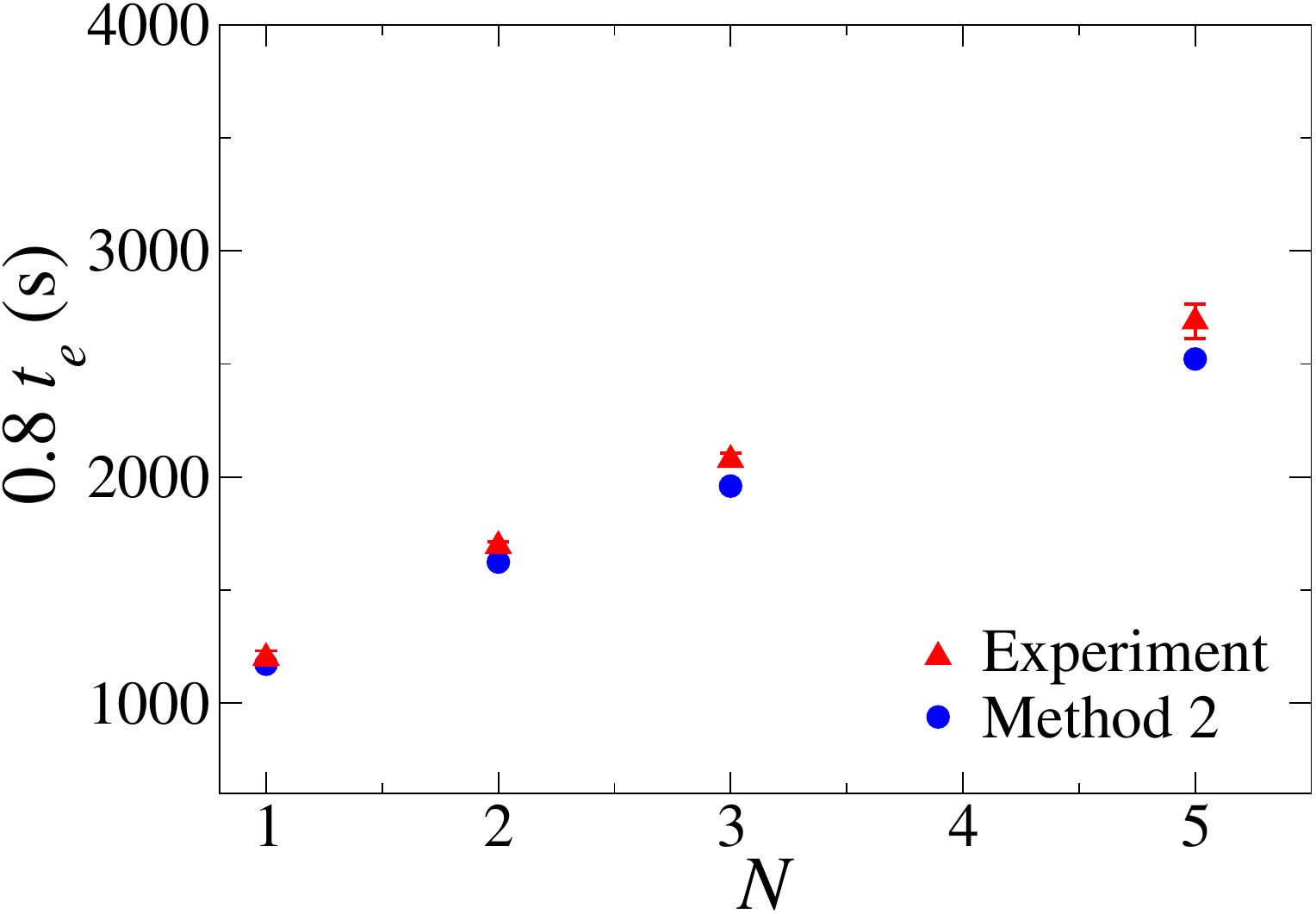}\\
\caption{Comparison of the 80\% lifetimes of the central droplet obtained experimentally with theoretical predictions as a function of the number of droplets ($N$) for (a,c) the partially pinned and (b,d) the completely pinned droplets. The first and second rows show predictions using averaged and exact variations of droplet parameters, respectively. The experimental data points are shown with error bars, representing the variability from three experimental repetitions.}
\label{fig:fig11}
\end{figure}

The general theory requires the instantaneous values of droplet radius ($R$), contact angle ($\theta$), and the separation distance between droplets ($D_{i,k}$). An exact application of the theory therefore necessitates time-resolved measurements of these variables from experiments. Fortunately, universal profiles of droplet diameter ($D$) and contact angle ($\theta$) can be obtained for both completely pinned and partially pinned configurations, as shown in Figures~\ref{fig:fig8} and \ref{fig:fig9}. These profiles are termed “universal” because they are independent of the number of droplets and depend solely on whether the dynamics is completely pinned or partially pinned. Consequently, only the initial droplet radius, contact angle, and $L/d$ value are needed to reconstruct the time-dependent variables for theoretical estimates using these profiles. In addition, the average ratio of side-to-central droplet lifetimes for each configuration (provided in Supplementary Table S6) is used. All droplets in a given configuration are assumed to have identical initial radius, contact angle, and volume at the onset of evaporation, with their subsequent evolution following the universal profiles. Theoretical predictions obtained using these averaged initial values and the universal profiles are represented as ``Method 1'', and are compared with experimental lifetimes in Figures~\ref{fig:fig11}(a,b). The agreement is satisfactory, with the maximum deviation between theory and experiment being approximately 12\% (Table \ref{table:T2}).

\begin{table}[]
\caption{Percentage deviation between experimental droplet lifetimes and theoretical predictions for partially pinned and completely pinned configurations.}
\label{table:T2}
\begin{tabular}{|c|cc|cc|}
\hline
\multirow{2}{*}{N} & \multicolumn{2}{c|}{Method 1}          & \multicolumn{2}{c|} {Method 2}          \\ \cline{2-5} 
                   & \multicolumn{1}{c|}{Partially pinned} & {Completely pinned} & \multicolumn{1}{c|}{Partially pinned} & {Completely pinned} \\ \hline
1                  & \multicolumn{1}{c|}{5.5}      & 3.4    & \multicolumn{1}{c|}{-3.3}     & 2.3    \\ \hline
2                  & \multicolumn{1}{c|}{11.5}     & 9.9    & \multicolumn{1}{c|}{3.6}      & 4.2    \\ \hline
3                  & \multicolumn{1}{c|}{9.5}      & 11.7   & \multicolumn{1}{c|}{7}        & 5.6    \\ \hline
5                  & \multicolumn{1}{c|}{-6.2}     & -10.3  & \multicolumn{1}{c|}{4.8}      & 6.2    \\ \hline
\end{tabular}
\end{table}

To address the residual deviations from theoretical predictions, we refined the initial value estimates by incorporating the individual droplet diameters and droplet-to-droplet distances for each configuration, rather than using averaged values. The individual initial values of $R$ and $D_{i,k}$ were extracted from top-view imaging of the evaporating drops, enabling us to capture experimental uncertainties in droplet radius and inter-droplet spacing. Assuming that all droplets had the same initial volume, we calculated the corresponding initial contact angles and evolved droplets with different initial radii and contact angles using the normalized universal profiles. Additionally, the exact ratio of each side-to-central droplet lifetime, determined from the top-view data, was incorporated into the model. The resulting predictions are labeled as “Method 2”. As shown in Figure \ref{fig:fig11}(c,d) and Table \ref{table:T2}, accounting for the initial variations and normalizing with individual droplet lifetimes significantly improves the agreement between theory and experiment. The maximum deviation is now only 7\%, and in most cases, the theoretical predictions lie within the experimental lifetime scatter. These results confirm that the analytical model described in Equation \ref{j:eq6} effectively captures the essential physics of multi-droplet evaporation at room temperature for both completely pinned and partially pinned cases. Further reductions in the remaining discrepancies could be achieved if the exact volumes of all droplets in the configurations were known.

\section{Conclusions}\label{sec:conclusions}

We conducted a systematic experimental study of sessile droplet evaporation in isolated and multi-droplet configurations on a substrate, focusing on the influence of contact line mobility. completely pinned droplets were produced by adding a low concentration of nanoparticles, sufficient to anchor the contact line without directly altering the evaporation rate, ensuring that any observed differences arose solely from the pinning effect. This approach allowed a direct comparison between completely pinned and partially pinned evaporation modes under identical ambient and geometric conditions. Configurations of one, two, three, and five droplets were examined at a fixed spacing ratio $(L/d=1.2)$ in controlled ambient environments. High-speed shadowgraphy was used to measure droplet height, wetted diameter, contact angle, and volume, while infrared thermography provided droplet temperature data for incorporating evaporative cooling into the analysis. Our results demonstrate that droplet lifetime increases systematically with the number of droplets due to vapor shielding, with a stronger relative enhancement for completely pinned droplets. In five-droplet configurations, lifetimes increased by 89\% for partially pinned and 124\% for completely pinned droplets compared to isolated droplets. The enhanced shielding in completely pinned droplets arises from the constant contact radius, which preserves close proximity between droplets throughout evaporation, thereby sustaining vapor–vapor interactions. In contrast, partially pinned droplets experience contact line recession, reducing proximity and diminishing the shielding effect over time. A diffusion-driven evaporation model, modified to include evaporative cooling, predicted central droplet lifetimes with good agreement to experimental measurements, with maximum deviations of about 11.7\%. These findings demonstrate the coupled influence of contact line dynamics and droplet arrangement on evaporation kinetics, providing insights for applications such as inkjet printing, spray cooling, biosensor fabrication, and coating technologies, where precise control of droplet lifetime and evaporation uniformity is essential. Future studies could extend this framework to non-uniform droplet arrays, larger temperature gradients, and forced convection to further explore the complex physics of evaporation in multidrop configurations.\\

\noindent{\bf Credit authorship contribution statement} 

Hari Govindha performed the experiments. All the authors contributed to the analysis of the results and to the preparation of the manuscript. The project was coordinated by Kirti Chandra Sahu.

\noindent{\bf Declaration of Competing Interest}

The authors declare that there is no conflict of interest.\\

\noindent{\bf Supplementary material}

\begin{itemize}
\item Table S1: Lifetime (in seconds) of droplets without and with 2.0 wt.\% nanoparticles on a cellulose acetate substrate.
\item Table S2: Best linear fit of droplet lifetime ($t_e = aN + b$) as a function of the number of droplets in the configuration ($N$).
\item Table S3: Parameters of the polynomial fits for normalized droplet height ($h$), wetted diameter ($D$), contact angle ($\theta$), and volume ($V$) for partially pinned droplets. 
\item Table S4: Parameters of the polynomial fits for normalized droplet height ($h$), wetted diameter ($D$), contact angle ($\theta$), and volume ($V$) for completely pinned droplets.
\item Properties of the working fluid at $24^\circ$C.
\item Table S6: Ratio of side-to-central droplet lifetimes in three- and five-droplet configurations.
\item Suppression matrix.

\end{itemize}



\providecommand{\latin}[1]{#1}
\makeatletter
\providecommand{\doi}
  {\begingroup\let\do\@makeother\dospecials
  \catcode`\{=1 \catcode`\}=2 \doi@aux}
\providecommand{\doi@aux}[1]{\endgroup\texttt{#1}}
\makeatother
\providecommand*\mcitethebibliography{\thebibliography}
\csname @ifundefined\endcsname{endmcitethebibliography}
  {\let\endmcitethebibliography\endthebibliography}{}

\end{document}